\documentclass[11pt]{article}
\usepackage{arxiv}
\usepackage{tikz}
\usetikzlibrary{arrows.meta, positioning, fit, backgrounds, calc}
\usepackage{pgfplots}
\pgfplotsset{compat=1.18}
\newcommand{\circlednum}[1]{\tikz[baseline=(c.base)]\node[circle, draw, inner sep=0.5pt, minimum size=3.4mm, font=\scriptsize\bfseries] (c) {#1};}

\newcommand{\brk}{\discretionary{}{}{}}
\title{Pre-Compiled Pipeline Shards for Distributed LLM Inference\\on Intel AI PC Fleets}

\author{
  Tate Berenbaum \\
  Not Community Labs Inc.
  \and
  Muthaiah Venkatachalam \\
  Intel Corporation
}

\date{August 2026}

\begin{document}

\maketitle

\begin{abstract}
Modern Intel AI PCs ship capable integrated GPUs and NPUs with 16+ GB of unified memory, and they spend considerable time idle. That is not enough memory to fit a large model such as a 70B-parameter LLM. We show that a handful of these machines, working together over an ordinary network, can serve models at and beyond the capability of any single one. Our system uses pipeline parallelism: a model is split by layer into per-stage shards, each pre-compiled into an INT4 OpenVINO graph, so that every machine runs one shard and passes activations to the next. Three techniques make this fast enough to be useful. First, we recover the speed of the unsplit model: a naive per-stage export runs well below monolithic inference because it misses an OpenVINO GPU optimization, and injecting a \texttt{beam\_idx} Gather into each shard triggers that optimization (the \texttt{Indirect\brk KVCache} fusion) and brings the shards to parity. Second, we make speculative decoding pay off on stateful OpenVINO models, which lack the paged-attention APIs that standard implementations assume: instead of physically trimming rejected draft tokens from the KV cache, we mask them out through the existing \texttt{attention\_\brk mask} input, which is bit-identical to a physical trim at essentially no cost. Third, the pipeline serves several users at once by interleaving their requests across the stages, each request carrying its own cache (micro-batching). Together, a two-node Llama~3.1~8B INT4 pipeline serves two concurrent users at $1.79\times$ the single-user throughput of the unsplit model on the same hardware, and the gap widens under simulated wide-area latency. The same design scales to a 70B model that no single fleet member can hold: a four-node deployment of Lunar Lake AI PCs on Intel Tiber Cloud serves a single user at interactive speed, with output token-for-token identical to the same four-node pipeline decoding without speculation. Code, raw benchmark logs, and reproduction scripts ship as a self-contained package at \url{https://github.com/labscommunity/pipeline-sharded-inference-paper} (in the top-level \texttt{reproduction/} directory).
\end{abstract}

\section{Introduction}
\label{sec:intro}

Modern AI PCs ship with integrated GPUs, NPUs, and 16+ GB of unified memory, and they spend considerable time idle. A single Intel AI PC runs Llama 3.1 8B~\cite{grattafiori2024llama} at a few tens of tokens per second. A fleet of them, coordinated over a standard office network, could serve LLM inference without cloud dependency---at zero marginal cost, with full data locality.

The barrier is software, not hardware. Standard model export tools (\texttt{torch.export}, \texttt{torch.\brk onnx.\brk export}) fail on modern transformer attention due to dynamic control flow in rotary embeddings and KV cache management. Monolithic compiled models cannot be split at layer boundaries without invasive graph surgery. Pipeline parallelism, developed for training throughput~\cite{huang2019gpipe, narayanan2019pipedream}, has seen limited evaluation for autoregressive \emph{inference} on consumer hardware---and when stages are distributed over a real network, each per-token TCP round trip becomes the dominant cost. Speculative decoding~\cite{leviathan2023fast, chen2023accelerating} is the natural answer (amortize network round-trips across multiple tokens emitted per target forward), but the canonical implementations rely on paged attention or dedicated rewind APIs that are not available on stateful OpenVINO models.

We build a distributed-inference stack that closes all three gaps, and the composition produces a surprisingly strong result: a two-node fleet of consumer Intel AI PCs serves two concurrent users at $1.79\times$ the throughput of a single-user monolithic baseline on the same hardware, and under a 100~ms/hop simulated WAN the same configuration stays usable while naïve pipeline-parallel decode falls below the interactive floor.

We make three contributions:

\begin{enumerate}[leftmargin=*]
  \item \textbf{A per-stage export pipeline that produces INT4 OpenVINO IR shards at monolithic parity.} Starting from \texttt{torch.\brk jit.\brk trace} with precomputed rotary embeddings~(\S\ref{sec:export}) and adding a post-export \texttt{beam\_idx} Parameter + \texttt{Gather(ReadValue, beam\_idx, axis=0)} injection that unlocks the OpenVINO GPU plugin's \texttt{Indirect\brk KVCache} transformation~(\S\ref{sec:shards}), the resulting shards reach within $4\%$ of \texttt{openvino\_genai.LLMPipeline} monolithic inference. Without the \texttt{beam\_idx} injection, the same shards run $13$--$23\%$ slower (and splitting into more stages widens the gap); the fusion unlock is the non-trivial part of the pipeline, and we document how we found it.

  \item \textbf{Mask-based KV-cache rewind for speculative decoding on stateful models.} Speculative decoding on an OpenVINO stateful target requires rejecting speculative draft tokens from the KV cache. The natural path---\texttt{query\_state()}/\texttt{set\_state()} physical trim---costs $\sim 48$~ms per call on Arc B390 iGPU with a 72-token cache (measured; one trim per step amortizing over the rejected drafts in that step), enough to make speculative decoding a net loss at break-even accept rates. Setting \texttt{attention\_\brk mask[j]\,=\,0} for rejected-draft cache positions is bit-exact with physical trim at essentially zero cost and enables $1.33\times$ mean single-node speedup across eight prompt types, rising to ${\sim}1.6\times$ at 2048-token generations~(\S\ref{sec:spec}). We document this as Discovery~\#20 in the accompanying repository.

  \item \textbf{Pipeline micro-batching via independent stateful InferRequests, composing multiplicatively into a distributed stack that beats monolithic single-user throughput and degrades gracefully under WAN latency.} Each InferRequest carries its own KV cache, so interleaving decode steps from separate user requests across pipeline stages yields $1.80\times$ system-throughput scaling on our $v_5$\_beam Llama~8B shards, with streams isolated via per-stream \texttt{compile\_\brk model}~(\S\ref{sec:microbatch}). Stacked on the first two contributions, the composition is empirical: the two-node fleet reaches $43.97$~tok/s at LAN ($1.79\times$ monolithic single-user), remains interactive at $100$~ms/hop simulated WAN, and scales to $64.67$~tok/s at $3$-stream on a three-node testbed~(\S\ref{sec:distributed}). A 4-stage Llama~3.1~70B INT4 deployment on Intel Tiber Cloud demonstrates the pipeline extends to model sizes that do not fit on any single fleet member~(\S\ref{sec:llama70b}). Top-1 logits compression recovers interactive throughput on relay-mediated WAN paths~(\S\ref{sec:topk_compress}).
\end{enumerate}

The system is evaluated on three Intel AI PCs (two ASUS Zenbook~S~14 with Lunar Lake, one HP OmniBook~X~16 with Panther Lake) connected over WiFi. Results are reproducible: every table row below is backed by a committed benchmark script and raw log under the \texttt{reproduction/} directory in the companion repository (see Section~\ref{sec:conclusion}).

\section{Background and Related Work}
\label{sec:related}

\subsection{Pipeline Parallelism for Inference}

Pipeline parallelism partitions a model's layers across devices. GPipe~\cite{huang2019gpipe} and PipeDream~\cite{narayanan2019pipedream} established this for training, where micro-batches keep all stages busy. Autoregressive inference is different: each token depends on the previous, creating strict sequential dependencies during decode.

Petals~\cite{borzunov2023petals} demonstrated decentralized pipeline inference across volunteer GPUs over the internet, running BLOOM-176B at roughly one decode step per second via DHT-based routing. Parallax~\cite{tong2025parallax} schedules pipelined LLM inference over decentralized, geographically separated consumer GPUs, cutting p99 latency by up to $2.6\times$ against decentralized-serving baselines. MDI-LLM~\cite{macario2025mdillm} proposed recurrent pipeline parallelism for edge devices, raising generation rate with each added node on a three-board Jetson TX2 testbed.

Our work differs in three respects. We pre-compile OpenVINO IR shards with a \texttt{beam\_idx}-Gather graph-surgery pass that unlocks the \texttt{Indirect\brk KVCache} GPU fusion, producing per-stage shards at monolithic parity rather than partitioning models at runtime with graph overhead. We target Intel AI PC integrated GPUs via OpenVINO rather than NVIDIA or Apple hardware. And we introduce mask-based speculative decoding rewind~(\S\ref{sec:spec})---a technique that makes speculative decoding~\cite{leviathan2023fast, chen2023accelerating} pay off on stateful OpenVINO without paged attention---which, stacked with micro-batching, produces the multi-user throughput advantage the paper builds toward.

We do not benchmark against Petals, Parallax, or other distributed inference systems because the hardware targets do not overlap: Petals and Parallax run on NVIDIA GPUs, MDI-LLM on Jetson boards, and our system on Intel iGPUs via OpenVINO. A cross-system comparison would conflate hardware differences with software differences. Instead, we compare against the monolithic \texttt{openvino\_genai.LLMPipeline} on the same hardware---the strongest single-node baseline available on our target platform---and report absolute throughput throughout so that readers with access to other systems can make their own comparisons.

\subsection{Model Compilation and Export}

The standard path from PyTorch to a compiled inference graph relies on abstract tensor shape propagation (\texttt{torch.export}, \texttt{torch.\brk onnx.\brk export}), which fails on the reshape and transpose operations inside modern rotary position embeddings. OpenVINO's \texttt{ov.\brk convert\_model} inherits these failures when given an \texttt{nn.Module}.

The \texttt{optimum-\brk intel} library exports whole models via \texttt{OVModelForCausalLM} but produces monolithic graphs with no per-layer splitting mechanism. KTransformers~\cite{chen2025ktransformers} compiles individual MoE experts for CPU/GPU hybrid inference but does not handle per-layer-range export for dense models.

Our pipeline uses \texttt{torch.\brk jit.\brk trace} with real tensors---avoiding abstract shape propagation entirely---and precomputes rotary embeddings externally, passing them as flat inputs. This sidesteps the tracing failures while producing per-stage graphs that the OpenVINO compiler optimizes independently.

\subsection{Serving Systems and Throughput Optimization}

vLLM~\cite{kwon2023vllm} introduced PagedAttention for GPU KV cache management. Sarathi-Serve~\cite{agrawal2024sarathi} proposed chunked prefills with stall-free scheduling. TD-Pipe~\cite{zhang2025tdpipe} temporally disaggregates prefill and decode phases, achieving up to 1.91$\times$ throughput. Splitwise~\cite{patel2024splitwise} splits these phases across specialized hardware.

These systems target data-center GPU clusters. Our micro-batching is simpler: each OpenVINO InferRequest carries independent KV cache state via ReadValue/Assign ops in the compiled graph, so temporal interleaving of requests requires no framework changes and no memory management beyond what OpenVINO already provides. The runtime also supports continuous batching~\cite{yu2022orca}, which the results tables do not use: OpenVINO's \texttt{ContinuousBatchingPipeline} provides it on CPU and GPU, and a packed multi-slot scheme of our own provides it on the NPU, whose compiler accepts no batched graph at all (\S\ref{sec:packed}).

A complementary line of work shrinks the KV cache itself rather than scheduling around it. GEAR~\cite{kang2024gear} compresses cache entries to 4~bits near-losslessly by pairing uniform quantization with a low-rank approximation of the residual and a sparse correction for outliers; SkipKV~\cite{tian2025skipkv} skips KV generation and storage outright for semantically redundant spans of long reasoning traces. Cache-footprint reduction of this kind is orthogonal to sharding and would compose with it, since each stage owns only its own layers' cache: KV growth is what erodes long-context throughput on our 70B deployment (\S\ref{sec:llama70b}), and per-slot cache capacity is the binding resource in the packed NPU serving mode (\S\ref{sec:packed}). We do not evaluate either technique; the one cache-precision knob we did test---OpenVINO's uniform INT8 KV cache hint---ran slower than the default on Arc iGPU (\S\ref{sec:negative}).

\section{System Design}
\label{sec:design}

\subsection{Overview}
\label{sec:overview}

\begin{figure}[ht]
\centering
\resizebox{\linewidth}{!}{%
\begin{tikzpicture}[
  font=\small,
  comp/.style={draw, rounded corners, fill=blue!8, minimum width=33mm, minimum height=8mm, inner sep=3pt, align=center},
  gpu/.style={draw, rounded corners, fill=orange!20, minimum width=33mm, minimum height=8mm, inner sep=3pt, align=center},
  nodebox/.style={draw, thick, rounded corners, inner sep=6pt},
  flow/.style={-{Stealth[length=2.5mm]}, very thick, blue!50!black},
  ret/.style={-{Stealth[length=2.5mm]}, very thick, dashed, red!60!black},
  wire/.style={font=\scriptsize, align=center},
  stepnum/.style={circle, draw, fill=white, inner sep=1pt, minimum size=4mm, font=\scriptsize\bfseries}
]
\node[comp] (loop) {generation loop \& tokenizer\\[-1pt]{\scriptsize spec-decode control, masks, \texttt{position\_\brk ids}}};
\node[gpu, below=4.5mm of loop] (s0) {stage-0 shard (iGPU)\\[-1pt]{\scriptsize embeddings $+$ layers $0..k$}};
\node[gpu, below=4.5mm of s0] (draft) {draft model (iGPU)\\[-1pt]{\scriptsize proposes $K$ tokens/step}};
\begin{scope}[on background layer]
\node[nodebox, fill=gray!8, fit=(loop)(s0)(draft)] (coord) {};
\end{scope}
\node[font=\small\bfseries, anchor=south] at (coord.north) {coordinator = fleet node 0};

\node[gpu, right=33mm of s0] (s1) {stage-1 shard (iGPU)\\[-1pt]{\scriptsize layers $k\!+\!1..m$}};
\begin{scope}[on background layer]
\node[nodebox, fill=gray!8, fit=(s1)] (w1) {};
\end{scope}
\node[font=\small\bfseries, anchor=south] at (w1.north) {worker node 1};

\node[gpu, right=16mm of s1] (sn) {final shard (iGPU)\\[-1pt]{\scriptsize layers $m\!+\!1..L$ $+$ \texttt{lm\_head}}};
\begin{scope}[on background layer]
\node[nodebox, fill=gray!8, fit=(sn)] (wn) {};
\end{scope}
\node[font=\small\bfseries, anchor=south] at (wn.north) {worker node $N\!-\!1$};

\draw[flow] (loop) -- node[stepnum, right=0mm] {1} (s0);
\draw[flow, <->] (loop.west) to[bend right=40] node[stepnum, left=0mm] {2} (draft.west);

\draw[flow] (coord.east |- s0) -- node[wire, below=1.5mm] {hidden state\\[-2pt]{\tiny \texttt{[1,n,4096]} fp32}} node[stepnum, above=1mm] {3} (w1.west |- s0);
\draw[flow] (w1.east |- s1) -- node[stepnum, above=0.5mm] {4} (wn.west |- s1);

\draw[ret] (wn.south) to[bend left=10] node[wire, below=2mm, pos=0.55] {token id (or top-1 logits), relayed hop-by-hop} node[stepnum, above=1mm, pos=0.45] {5} (coord.south east);
\end{tikzpicture}}
\caption{System architecture and per-step decode flow. \emph{Offline}, the export pipeline (\S\ref{sec:export}) turns a HuggingFace checkpoint into $N$ self-contained INT4 OpenVINO IR shards, one per node. \emph{Online}, each decode step proceeds: \protect\circlednum{1} the coordinator runs the stage-0 shard on its local iGPU; \protect\circlednum{2} under speculative decoding (\S\ref{sec:spec}) the local draft model proposes $K$ tokens, so the pipeline verifies $K\!+\!1$ positions per traversal; \protect\circlednum{3} the hidden state crosses to the next stage over a persistent TCP connection (\S\ref{sec:relay}), $\sim$16~KB per token per hop; \protect\circlednum{4} intermediate stages forward downstream; \protect\circlednum{5} the final stage applies the LM head and the sampled token (or top-1 logits, \S\ref{sec:topk_compress}) is relayed hop-by-hop back to the coordinator, which appends it and repeats. Under micro-batching (\S\ref{sec:microbatch}), 2--3 user streams interleave through the same chain, each with its own KV state on every stage.}
\label{fig:system}
\end{figure}
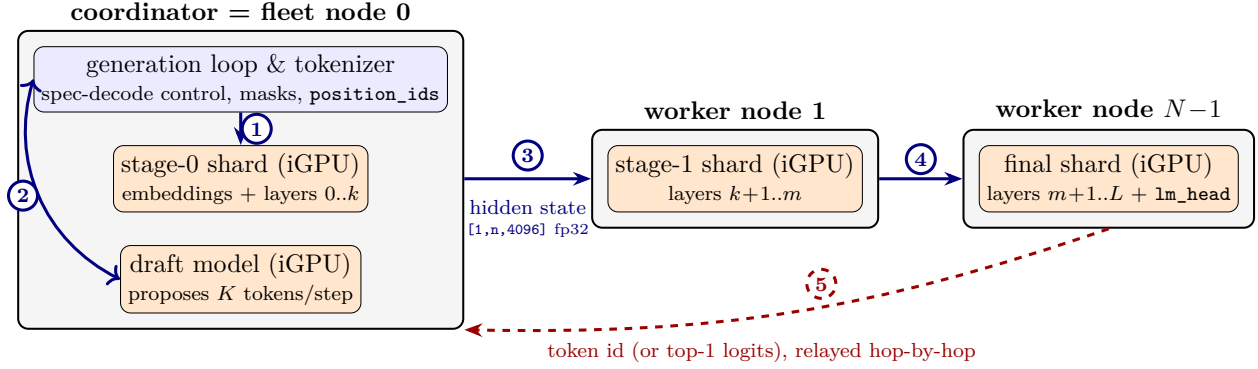

Four components: (1)~a per-stage export pipeline producing compiled INT4 OpenVINO~\cite{openvino2026} IR shards with stateful KV cache, (2)~a TCP activation relay, (3)~stage workers that load and run their assigned shard, and (4)~a coordinator driving the autoregressive generation loop. Figure~\ref{fig:system} shows how the components fit together and how a decode step flows through them.

\textbf{Coordinator placement.} The coordinator is not a separate orchestration machine: it is itself a fleet member (node~0 in Figure~\ref{fig:system}) hosting the tokenizer, the generation loop, the stage-0 shard, and---when speculative decoding is enabled---the draft model, the latter two on its local iGPU. Every multi-node result in this paper runs the coordinator this way. The stage-0 device is configurable: a CPU-hosted stage~0 works and is measured for Gemma~4 (Table~\ref{tab:gemma_dist}, CPU$\rightarrow$GPU at $9.87$~tok/s vs.\ $10.40$ GPU$\rightarrow$GPU), but all headline configurations place every stage, draft model included, on integrated GPUs.

Each worker node loads one shard and exposes a TCP endpoint. Per decode step, the coordinator runs stage~0 locally and sends the resulting hidden state to the next node; each intermediate stage forwards its output downstream; the final stage applies the LM head and returns the next token, which is relayed hop-by-hop back to the coordinator.

\subsection{Per-Stage Export Pipeline}
\label{sec:export}

The pipeline converts a HuggingFace transformer into $N$ standalone OpenVINO IR shards, each covering a contiguous layer range plus (for the first) the embedding and (for the last) the output head.

\textbf{Step 1: Selective layer loading.} Decoder layers are loaded from safetensors via a model-agnostic structure mapping (Llama, Mistral, Qwen, Phi, Gemma). Each stage gets only its assigned layers. Memory usage scales with layer count, not model size.

\textbf{Step 2: Attention rewrite.} HuggingFace's \texttt{DynamicCache} (Python list ops) and \texttt{Llama\brk Rotary\brk Embedding} (\texttt{torch.autocast} guards) are both untraceable. We replace them with a manual attention implementation taking precomputed cos/sin tensors and explicit KV tensors as inputs. Numerical equivalence verified: max difference $< 5 \times 10^{-7}$ against HuggingFace's native forward.

\textbf{Step 3: Trace and convert.} Each stage is traced with \texttt{torch.\brk jit.\brk trace} using real tensors (batch size~1, representative sequence length). The traced graph converts to OpenVINO IR via \texttt{ov.\brk convert\_model}. KV cache tensors become stateful ReadValue/Assign pairs via \texttt{make\_\brk stateful\_\brk transformation}.

\textbf{Step 4: INT4 compression.} INT4 symmetric weight compression via NNCF~\cite{nncf2024}, group size~128. Each shard is a self-contained IR (XML~+~BIN) taking token IDs or hidden states in, producing hidden states or logits out.

\subsection{TCP Activation Relay}
\label{sec:relay}

Activations travel between nodes over plain operating-system TCP sockets. There is no peer-to-peer networking framework, DHT, or relay infrastructure in the path: decentralized systems such as Petals~\cite{borzunov2023petals} route stages through libp2p, which negotiates among multiple transports, but a fleet under one administrative domain needs neither peer discovery nor NAT traversal nor transport negotiation, and every layer removed from the per-token path matters when the whole token budget is ${\sim}60$~ms.

\textbf{Connection establishment.} Each worker listens on a configured \texttt{host:port}. At startup the coordinator dials stage~1 and each stage~$i$ dials stage~$i\!+\!1$, retrying until the downstream worker is up so that bring-up order does not matter. The result is a static daisy chain of persistent TCP connections, one per adjacent stage pair, held open for the entire session: no per-token or per-request connection setup, and no handshakes on the decode path. Tokens emitted by the final stage return to the coordinator hop-by-hop over the same sockets (Figure~\ref{fig:system}).

\textbf{Wire format.} Each message is a 20-byte header (payload length, dtype, three shape dimensions) followed by raw tensor bytes. KV-cached decode sends \texttt{[1,\,1,\,4096]} float32 tensors: 16~KB per hop. \texttt{TCP\_NODELAY} is set on every socket so the kernel does not buffer the small per-token writes (Nagle's algorithm would otherwise delay them by up to one round trip). The protocol is intentionally minimal---designed to be swapped for QUIC, RDMA, compression, or other transports without touching the rest of the system; \S\ref{sec:compression} discusses what a transport swap could and could not buy.

\subsection{Stateful KV-Cached Inference}

Each shard's compiled graph contains ReadValue/Assign ops implementing the KV cache. Prefill processes the full prompt in one pass; decode processes one token per step, reading cached context and appending the new KV pair. The coordinator precomputes rotary cos/sin tensors for the maximum sequence length and slices the appropriate position range per step.

For models with cross-layer KV sharing (Gemma~4 shares K/V between layers via \texttt{DynamicCache}), the shared tensors are transmitted as additional stage outputs/inputs over TCP, with a 4D$\rightarrow$3D reshape for the 3-dimension header protocol.

\subsection{Operating Scenarios and Contribution Map}
\label{sec:scenarios}

The runtime composes into three operating scenarios, in increasing order of machinery:

\begin{enumerate}[label=(\alph*), leftmargin=*]
  \item \textbf{PP}: pipeline-parallel decode alone---one user, one token per traversal of the stage chain.
  \item \textbf{PP$+$SD}: pipeline parallelism with speculative decoding---the coordinator-local draft model proposes $K$ tokens and the pipeline verifies all $K\!+\!1$ positions in a single traversal~(\S\ref{sec:spec}).
  \item \textbf{PP$+$SD$+$MB}: scenario~(b) plus multi-user micro-batching---multiple concurrent user streams interleaved across the same stages, each with independent KV state~(\S\ref{sec:microbatch}).
\end{enumerate}

\begin{table}[ht]
\centering
\caption{Mapping the three contributions of \S\ref{sec:intro} to the three operating scenarios. Parenthesized marks on contribution~3: only its composition-and-WAN component applies in scenarios~(a) and~(b); micro-batching itself enters in scenario~(c).}
\label{tab:contrib_map}
\begin{tabular}{@{}lccc@{}}
\toprule
Contribution & (a) PP & (b) PP$+$SD & (c) PP$+$SD$+$MB \\
\midrule
1. Shard export $+$ \texttt{beam\_idx} fusion unlock (\S\ref{sec:export}, \S\ref{sec:shards}) & \checkmark & \checkmark & \checkmark \\
2. Mask-based KV rewind (\S\ref{sec:spec}) & --- & \checkmark & \checkmark \\
3. Micro-batching $+$ composition / WAN (\S\ref{sec:microbatch}, \S\ref{sec:distributed}) & (\checkmark) & (\checkmark) & \checkmark \\
\bottomrule
\end{tabular}
\end{table}

Contribution~1 (the export pipeline and the \texttt{beam\_idx} fusion unlock) underlies all three scenarios: it produces the shards everything else runs on. Contribution~2 (mask-based KV rewind) is what makes speculative decoding affordable on stateful OpenVINO, so it enters in scenarios~(b) and~(c). Contribution~3 has two components. Micro-batching itself enters in scenario~(c). The composition-and-WAN component spans all three, since it is the empirical demonstration that (a)$\rightarrow$(b)$\rightarrow$(c) compose multiplicatively, plus the behavior of each scenario under WAN latency. \S\ref{sec:scenario_compare} compares the three scenarios head-to-head on the same topologies and discusses which to deploy when.

\textbf{Beam search is not an operating mode.} All decoding in this paper is greedy ($\arg\max$); we never run beam search. The \texttt{beam\_idx} input of \S\ref{sec:shards} exists purely as a compile-time graph pattern: the OpenVINO GPU plugin's \texttt{Indirect\brk KVCache} fusion fires only when each KV \texttt{ReadValue} flows through a \texttt{Gather} indexed by a \texttt{beam\_idx} Parameter. At runtime every stage receives the constant \texttt{beam\_idx}~$=[0]$, making the Gather an identity reorder of the batch-1 cache. In particular, there is no interaction between \texttt{beam\_idx} and the draft model in scenarios~(b)/(c): cache reordering (constant identity, driven by \texttt{beam\_idx}) and draft-rejection rewind (driven by \texttt{attention\_\brk mask}, \S\ref{sec:spec}) act on disjoint model inputs, so no \texttt{beam\_idx} conflict can arise. Running \emph{actual} beam search on top of speculative decoding would require per-beam draft state and a candidate-tree verify step, which we have not built; the exported graphs already contain the cache-reorder machinery, so the shard side is ready for it.

\section{Reaching Monolithic Parity: the beam\_idx Gather Injection}
\label{sec:shards}

A naive reading of the export pipeline of \S\ref{sec:export} gives a workable but uncompetitive result: on Llama~3.1~8B INT4, a 1-stage shard (all 32~layers in one graph, same weights as the monolithic model, only the export path differs) runs $13$--$23\%$ slower than \texttt{openvino\_genai.LLMPipeline} at steady state. Splitting into more stages makes it worse. The reason is that the OpenVINO GPU plugin contains a plugin-internal transformation, \texttt{Indirect\brk KVCache}, that rewrites \texttt{ReadValue $\rightarrow$ Concat $\rightarrow$ Assign} patterns into a fused \texttt{IndirectSDPA} op---but the transformation is gated on the presence of a specific pattern: each KV \texttt{ReadValue}'s output must flow through a \texttt{Gather} indexed by a \texttt{beam\_idx} Parameter (used by beam search in the monolithic \texttt{LLMPipeline} path). Our initial exports did not include that pattern, so the GPU plugin silently fell back to a generic \texttt{KVCache\brk Fusion} path with measurably lower XVE occupancy.

The fix is post-export graph surgery: after \texttt{apply\_\brk make\_\brk stateful\_\brk transformation}, walk the graph, add a \texttt{beam\_idx: [-1] i32} Parameter, and for each KV \texttt{ReadValue} insert a \texttt{Gather(ReadValue, beam\_idx, axis=0)} between the ReadValue output and its downstream consumers. This is exactly what \texttt{optimum-\brk intel}'s internal \texttt{fuse\_\brk cache\_\brk reorder} pass does for monolithic exports, lifted out and applied to our per-stage exports. We call the resulting shards $v_5$\_beam. To be explicit: the injection is a compile-time pattern only. At runtime every stage receives the constant \texttt{beam\_idx}~$=[0]$, the Gather is an identity reorder, and no beam search is performed anywhere in this paper (\S\ref{sec:scenarios}).

\subsection{Results}

\begin{table}[ht]
\centering
\caption{Single-node Llama~3.1~8B INT4 on Panther Lake Arc B390 iGPU, $N\!=\!5$ paired runs, same session, controlled-tokenization 128-token decode. $v_5$\_beam closes the export-to-monolithic gap from $-13$ to within $0.4\%$ of \texttt{openvino\_genai}. Re-measured 2026-04-28 on OV~2026.1.0; the original 2026-04-24 paired session showed $A=22.96$ tok/s.}
\label{tab:shard_parity}
\begin{tabular}{@{}lrrr@{}}
\toprule
Configuration & Tok/s & Range & vs.\ A \\
\midrule
A: mono \texttt{openvino\_genai} (C++ loop)       & 24.54 & 23.95--25.06 & 1.00$\times$ \\
A$'$: mono \texttt{ov.compile\_\brk model} (our Python loop) & 23.12 & 22.59--23.55 & 0.94$\times$ \\
B ($v_5$\_beam 1-stage, our export)               & 24.45 & 24.08--24.81 & 1.00$\times$ \\
B$_{v_3\_fp32}$ (pre-\texttt{beam\_idx})          & 21.26 & 20.93--21.55 & 0.87$\times$ \\
\bottomrule
\end{tabular}
\end{table}

The \texttt{beam\_idx} Gather injection recovers $15\%$ of throughput (from $21.26$ to $24.45$~tok/s) without touching the weights, tokenization, or decode loop. The remaining gap between B ($v_5$\_beam) and A has effectively closed (within $0.4\%$ on the same paired session): \texttt{openvino\_genai}'s C++ decode loop vs.\ our Python loop now shows a $\sim$$6\%$ Python overhead (A$'$~vs.\ A; was $\sim$$10\%$ on the prior measurement window), and B is $5.7\%$ \emph{faster} than A$'$ on the same paired measurement.

\subsection{Splitting into More Stages}

\begin{table}[ht]
\centering
\caption{$v_5$\_beam shards, 2-stage and 3-stage, in-process on Panther Lake iGPU. Same in-process baseline as the spec-decode composition table (Table~\ref{tab:spec_shard}): $A_{\text{spec}}=24.28$, slightly below Table~\ref{tab:shard_parity}'s $A=24.54$ due to a different paired session and run-to-run variance. The parity extends through splitting at the cost of ${\sim}1$~tok/s per additional stage from Python activation passing and smaller per-stage XVE problem sizes.}
\label{tab:shard_stages}
\begin{tabular}{@{}lrr@{}}
\toprule
Configuration & Tok/s & vs.\ $A_{\text{spec}}$ (24.28) \\
\midrule
Mono ($A_{\text{spec}}$, reference) & 24.28 & 1.00$\times$ \\
$v_5$\_beam, 1-stage             & 23.93 & 0.99$\times$ \\
$v_5$\_beam, 2-stage (16+16)     & 20.56 & 0.85$\times$ \\
$v_5$\_beam, 3-stage (11+11+10)  & 21.56 & 0.89$\times$ \\
\bottomrule
\end{tabular}
\end{table}

Per-stage splitting costs ${\sim}11$--$15\%$ on a single machine relative to $A_{\text{spec}}=24.28$ (1-stage is essentially at parity, 2-stage at $15\%$ overhead, 3-stage at $11\%$); the per-stage GEMMs are smaller, reducing XVE occupancy (measured $60.1\%$ on mono vs.\ $43.6\%$ on 3-stage via VTune \texttt{gpu-hotspots}, with identical top-kernel counts), and the Python-level \texttt{hidden\_states.astype(np.float32)} cast between stages is not free. This gap is structural: smaller per-stage problems cannot saturate the GPU the way a single large problem does. We return to how to recover this cost (and then some) via multi-user micro-batching in~\S\ref{sec:microbatch} and via speculative decoding in~\S\ref{sec:spec}.

\subsection{Methodology}

The numbers in Table~\ref{tab:shard_parity} come from one paired session using identical tokenization (chat template applied to both paths so \texttt{input\_ids} are byte-identical), identical \texttt{max\_new\_tokens}~=~128 with \texttt{ignore\_eos=True} (so EOS-driven early termination cannot bias short runs), and identical decode loops except where noted (A vs.\ A$'$ isolates loop language, A$'$ vs.\ B isolates the export path). $N\!=\!5$ runs after 2 warmups. Replication commits, raw JSON, and VTune summary reports are in \texttt{reproduction/} (see Section~\ref{sec:conclusion}).

\section{Speculative Decoding via Mask-Based KV Rewind}
\label{sec:spec}

Speculative decoding~\cite{leviathan2023fast, chen2023accelerating} amortizes target-model forward passes over multiple emitted tokens: a smaller draft model proposes $K$ tokens, the target verifies all $K$ in one forward pass, and accepted prefixes are emitted together. On a distributed pipeline this is doubly attractive---it amortizes not only target compute but also network round-trips. On a \emph{single} node, the technique's per-step cost structure determines whether it pays off at all.

\subsection{The Trim Problem}

Standard speculative decoding rejects draft tokens whose predictions do not match the target's. On an OpenVINO stateful model, rejecting tokens means removing their K and V contributions from the KV cache before the next target forward. The natural API for this is \texttt{Infer\brk Request.\brk query\_state()} (returns a list of the per-layer state tensors) followed by assigning a sliced tensor back to each state via \texttt{sv.state = ov.Tensor(sliced\_np)}. We measured this on Llama~3.1~8B INT4 on Arc B390:

\begin{table}[ht]
\centering
\caption{Per-call cost of physical KV trim on Arc B390 iGPU with a 72-token cache (32 layers, 64 state tensors). All three variants cluster around $45$--$50$~ms; the cost is device-side state invalidation, not numpy work.}
\label{tab:trim_cost}
\begin{tabular}{@{}lr@{}}
\toprule
\texttt{trim\_kv} variant & ms/call \\
\midrule
\texttt{np.asarray(sv.state.data)} + slice + new Tensor & 49.5 \\
In-place \texttt{Tensor.set\_shape()} & 46.6 \\
\texttt{ov.Tensor} allocate + \texttt{np.asarray(new)[:]} & 49.9 \\
\bottomrule
\end{tabular}
\end{table}

A $K\!=\!3$ speculative step executes one target verify ($\sim 47$~ms memory-bound forward of $K\!+\!1\!=\!4$ tokens), 2--3 draft feeds ($\sim 16$~ms each), and one correction feed ($\sim 16$~ms). At roughly 3 emitted tokens per step, the $\sim 48$~ms trim per step (Table~\ref{tab:trim_cost}) is enough to push the whole technique below break-even.

\subsection{Mask-Based Rewind}

Our alternative avoids state manipulation entirely. The OpenVINO stateful model takes \texttt{attention\_\brk mask} as a per-forward input of shape $[\mathrm{batch},\,\mathrm{past\_len}+\mathrm{input\_len}]$, where \texttt{mask[j]=0} means ``ignore cache position $j$ in attention.'' If rejected-draft tokens remain physically in the cache but their positions are masked out on every subsequent forward, the attention result is identical to a physically-trimmed cache. Future tokens' key and value tensors append at the next cache index, with RoPE computed from the caller-supplied \texttt{position\_\brk ids}; as long as \texttt{position\_\brk ids} tracks the \emph{logical} sequence length (the pre-rewind count) rather than the physical cache length, the rotary rotations are coherent.

We verified this is bit-exact with physical trim. Running both paths side-by-side on the same target--prompt--draft--correction sequence (\texttt{reproduction/\brk scripts/\brk bench/\brk mask\_trim\_test.py}), the maximum absolute difference in post-correction logits was $0.0000$ (fp32 equality to machine precision).

The cost is CPU-only: each forward reconstructs the $[1, \mathrm{past\_len}+\mathrm{input\_len}]$ mask by concatenating a tracked \texttt{valid\_mask} prefix with ones for the new tokens. Measured: $<\,1\%$ of per-step wall time at $K\!=\!3$. The memory cost is that rejected-draft K and V entries persist in cache, inflating cache size over generations. At 2048-token generation with $K\!=\!3$, cache bloat reaches $1.02\times$ logical length (most rejected-draft overhead is amortized out by the live-token count), so no periodic compaction is needed for realistic single-turn lengths. For multi-turn conversations with cumulative context approaching the model's window, periodic compaction (physically trimming the cache to remove all masked positions) would be needed; masked positions are compacted away when cache state is serialized at end of generation (\S\ref{sec:prefix_cache}), but in-generation compaction remains unimplemented and unmeasured. The mask primitive is also not specific to draft rejection: cache-reduction policies that drop low-utility positions mid-generation---e.g., SkipKV~\cite{tian2025skipkv}, which skips storage for semantically redundant sentences of a reasoning trace---presuppose cheap removal of arbitrary cache positions, exactly what the $\sim 48$~ms state round-trip of Table~\ref{tab:trim_cost} denies a stateful OpenVINO model. Masking those positions out instead yields the same attention result at no device cost, so the machinery built here would carry such policies as well.

\textbf{Scope: greedy decode only.} All results in this paper use greedy decoding ($\arg\max$ token selection). The mask-based rewind is correct for greedy because the acceptance check is a simple token-id comparison and the surviving cache is identical regardless of rewind method. Extending to temperature-based sampling would require the rejection-sampling correction of~\citet{leviathan2023fast} and~\citet{chen2023accelerating}, which adjusts the target distribution conditioned on draft probabilities; the mask-based rewind composes with this correction in principle (the cache state is the same either way), but we have not validated it empirically. Beam search is likewise not used anywhere in this paper; \S\ref{sec:scenarios} explains why the \texttt{beam\_idx} input nonetheless appears in every shard graph.

\subsection{Single-Node Results}

\begin{table}[ht]
\centering
\caption{Speculative decoding on Llama~3.1~8B INT4 (target) + Llama~3.2~1B INT4 (draft) on Arc B390 iGPU, greedy decode, bit-exact output vs.\ target-only greedy. Eight diverse prompts, 128-token decode, $K\!=\!3$, mean $\pm$ stddev.}
\label{tab:spec_lan}
\begin{tabular}{@{}lrrr@{}}
\toprule
Prompt category & Baseline (tok/s) & Spec (tok/s) & Speedup \\
\midrule
short-factual    & 22.60 & 29.35 & 1.30$\times$ \\
reasoning        & 21.47 & 29.48 & 1.37$\times$ \\
code-completion  & 22.08 & 33.10 & 1.50$\times$ \\
list-enumeration & 21.37 & 27.91 & 1.31$\times$ \\
creative         & 23.51 & 26.09 & 1.11$\times$ \\
technical-expl   & 23.87 & 28.52 & 1.19$\times$ \\
chat-assistant   & 23.30 & 31.22 & 1.34$\times$ \\
translation      & 23.04 & 34.14 & 1.48$\times$ \\
\midrule
\textbf{Mean}    & \textbf{22.66} & \textbf{29.98} & \textbf{1.33$\times$ $\pm$ 0.13} \\
\bottomrule
\end{tabular}
\end{table}

Speedup varies from $1.11\times$ (creative writing, 49.7\% accept rate) to $1.50\times$ (code completion, 93.1\% accept rate). On content that is easier for the draft to predict (structured output: code, translation, chat boilerplate), the speedup climbs toward the theoretical $1\!+\!K\!\cdot\!p_\text{accept}$ ceiling; on content where the draft diverges early, it falls to the acceptance floor. Every prompt produces output that is bit-exact identical to the target's own greedy decode on the first ten generated tokens (\texttt{reproduction/}'s smoke test verifies full-sequence equality).

Generation length amplifies the effect: across 128, 512, 1024, and 2048-token runs on the \emph{same} prompt, acceptance rises from $70.7\%$ to $97.5\%$ as the model enters its repetition regime, and speedup at $K\!=\!3$ climbs from $\sim 1.3\times$ at 128 tokens to $\sim 1.6\times$ at 2048 tokens.

\subsection{Compose-with-Shards}

Because the rewind lives at the Python wrapper layer---not inside the compiled OV graph---the same \texttt{MaskedReq} abstraction composes with both the monolithic target and our $v_5$\_beam 3-stage sharded target. Measured in a single paired session (\texttt{reproduction/\brk scripts/\brk bench/\brk bench\_\brk spec\_\brk matrix.py}):

\begin{table}[ht]
\centering
\caption{Speculative decoding stacks cleanly with sharding. Single-session paired measurement, $K\!=\!3$, 128-token decode.}
\label{tab:spec_shard}
\begin{tabular}{@{}lrrr@{}}
\toprule
Target & Tok/s & vs.\ mono-$A_{\text{spec}}$ & vs.\ own baseline \\
\midrule
Mono ($A_{\text{spec}}$) & 24.28 & 1.00$\times$ & --- \\
3-stage $v_5$\_beam shard & 21.56 & 0.89$\times$ & --- \\
Mono + spec $K\!=\!3$ & 30.21 & 1.24$\times$ & 1.24$\times$ \\
3-stage shard + spec $K\!=\!3$ & 28.38 & 1.17$\times$ & 1.32$\times$ \\
\bottomrule
\end{tabular}
\end{table}

The spec-decode multiplier is comparable on both paths ($1.24\times$ on the monolithic target, $1.32\times$ on the 3-stage shard target). Output is bit-exact: the mono-plus-spec output matches the mono-only output, and the shard-plus-spec output matches the shard-only output.

\section{Distributed Pipeline Evaluation}
\label{sec:distributed}

\subsection{Testbed}
\label{sec:testbed}

Three Intel AI PCs on the same WiFi network (802.11ax):

\begin{itemize}[leftmargin=*]
  \item \textbf{node-00} (\emph{beta}, ASUS Zenbook S 14): Core Ultra 7 258V (Lunar Lake), Arc 140V iGPU, 32~GB LPDDR5X-8533.
  \item \textbf{node-01} (\emph{charlie}, ASUS Zenbook S 14): Core Ultra 7 258V (Lunar Lake), Arc 140V iGPU, 32~GB LPDDR5X-8533.
  \item \textbf{node-02} (\emph{alpha}, HP OmniBook X 16): Core Ultra X7 358H (Panther Lake), Arc B390 iGPU, 32~GB DDR5.
\end{itemize}

All run Windows~11, Python~3.11, OpenVINO~2026.1.0, GPU inference. Raw TCP round-trip for 16~KB payloads: 6.5~ms on the 802.11ax LAN (measured separately).

\subsection{Progressive Optimization}
\label{sec:fullstack}

\begin{table}[ht]
\centering
\caption{Llama~3.1~8B INT4 distributed throughput on the 2-node \emph{alpha} + \emph{charlie} $v_5$\_beam pipeline, building up from the single-stream distributed baseline. System throughput for multi-user is aggregated across concurrent streams. All configurations produce bit-exact output relative to target-only greedy. The full-stack number ($43.97$) is from one paired session of \texttt{reproduction/\brk scripts/\brk coord/\brk mini\_\brk coord\_\brk spec\_\brk mbatch.py} on OpenVINO~2026.1.0. Run-to-run variance is typically $\pm 0.3$ stddev across sessions.}
\label{tab:progression}
\begin{tabular}{@{}lrrr@{}}
\toprule
Configuration & Tok/s & vs.\ mono-$A$ & vs.\ single-stream dist. \\
\midrule
Mono single-node reference ($A$, Table~\ref{tab:shard_parity}) & 24.54 & 1.00$\times$ & --- \\
2-stage $v_5$\_beam, single stream (\emph{alpha}+\emph{charlie}) & 16.33 & 0.67$\times$ & 1.00$\times$ \\
\,+ 2-stream micro-batching (\S\ref{sec:microbatch}) & 29.34 & 1.20$\times$ & 1.80$\times$ \\
\,+ mask-based spec decode $K\!=\!3$ (\S\ref{sec:spec}) & \textbf{43.97} & \textbf{1.79$\times$} & \textbf{2.69$\times$} \\
\bottomrule
\end{tabular}
\end{table}

Starting from the 2-node $v_5$\_beam pipeline at $16.33$~tok/s ($20$~ms slower per token than the $24.54$~tok/s single-node mono reference; of that $20$~ms, ${\sim}6$~ms is one TCP round-trip and ${\sim}14$~ms is Python wrapper plus OpenVINO dispatch overhead, see Table~\ref{tab:breakdown}), two composable multipliers stack multiplicatively:

\textbf{Micro-batching ($1.80\times$ over single-stream distributed).} Two independent \texttt{InferRequest}s per shard let the coordinator interleave stage-0 compute for request~B while stage~1 is busy with request~A. Our $v_5$\_beam micro-batch multiplier ($29.34 / 16.33 = 1.80\times$) reflects the smaller per-stage compute window that \texttt{Indirect\brk KVCache} fusion leaves behind---more idle time for the other stream to fill (Table~\ref{tab:microbatch} compares against an earlier internal export iteration). The mbatch ratio compresses on faster single-stream baselines (the $1.80\times$ here vs.\ $2.03\times$ on the prior measurement window) because the more-saturated baseline leaves less stage-idle time for the second stream to absorb.

\textbf{Mask-based spec decode ($1.50\times$ on top of mbatch).} Each of the two streams runs its own speculative decoding loop (\S\ref{sec:spec}). The $K\!+\!1$-token target verify batches four tokens per TCP round-trip and amortizes the per-hop latency. Applied atop the two-stream pipeline, this yields a further $\sim 1.50\times$ multiplier, for a system-throughput total of $43.97$~tok/s: $1.79\times$ the monolithic single-user reference on the same hardware, two users served concurrently.

Composition is empirical, not a projection: $16.33 \times 1.797 \times 1.499 = 43.99$, matching the measured $43.97$ within $0.05$~tok/s; all three multipliers were measured in the same paired session (\texttt{reproduction/\brk scripts/\brk coord/\brk mini\_\brk coord\_\brk spec\_\brk mbatch.py}, reproduced 2026-04-28).

\subsection{Per-Token Breakdown}

\begin{table}[ht]
\centering
\caption{Where the time goes: 2-stage $v_5$\_beam Llama~3.1~8B distributed pipeline on the \emph{alpha}+\emph{charlie} 802.11ax WiFi LAN, GPU--GPU, single stream (no spec decode, no micro-batching). Compute numbers averaged across N=3 timed runs.}
\label{tab:breakdown}
\begin{tabular}{@{}lrr@{}}
\toprule
Component & Time & \% of token \\
\midrule
Stage~0 compute (coordinator, \emph{alpha} Panther Lake) & 20~ms & 33\% \\
Stage~1 compute (worker, \emph{charlie} Lunar Lake) & 22~ms & 36\% \\
TCP round trip (single hop, 802.11ax) & 6~ms & 10\% \\
Python activation passing + mask construction & 13~ms & 21\% \\
\midrule
\textbf{Total} & \textbf{61~ms} & $\rightarrow$ \textbf{16.33~tok/s} \\
\bottomrule
\end{tabular}
\end{table}

\textbf{Prefill latency (TTFT).} For interactive chat, time to first token matters as much as decode throughput. On the 2-stage distributed pipeline, TTFT ranges from $122$~ms (short 8-token prompt) to $786$~ms (long prompt requiring multi-pass prefill), dominated by prompt-length-dependent compute on stage~0 rather than network overhead. On the 4-stage 70B Tiber Cloud topology, we did not separately measure TTFT; the reported numbers are end-to-end decode throughput. Prefill latency characterization for the 70B case is future work.

An important subtlety: our initial instrumentation reported ``70\% network time,'' but that metric conflated actual TCP latency with the time spent waiting for the remote worker to compute. Dedicated TCP benchmarking measured the true network cost at ${\sim}6$~ms per round trip ($10\%$ of per-token time; the round-figure $6.5$~ms cited elsewhere is the same number rounded differently across sessions). The rest is compute and Python overhead (the latter is OpenVINO's C++ dispatch and state management, not user Python---a dedicated profile run (\texttt{reproduction/\brk scripts/\brk bench/\brk bench\_\brk feed\_\brk overhead.py}) finds user Python is $0.6\%$ of wall time).

\subsection{Activation Compression}
\label{sec:compression}

\begin{table}[ht]
\centering
\caption{Activation compression, 2-stage distributed Llama~3.1~8B. On LAN, bandwidth is not the bottleneck---compression cannot help.}
\label{tab:compression}
\begin{tabular}{@{}lrrrl@{}}
\toprule
Compression & Payload/hop & Tok/s & Decode & Quality \\
\midrule
None (FP32) & 16~KB & 13.57 & 65--68~ms & Correct \\
FP16 & 8~KB & 12.71 & 66--67~ms & Correct \\
INT8 & 4~KB & 15.10 & 63--68~ms & Corrupted \\
\bottomrule
\end{tabular}
\end{table}

At gigabit WiFi, 16~KB transmits in under $0.2$~ms. The $\sim 6.5$~ms per-hop latency is TCP kernel scheduling and round-trip time, not bandwidth. Halving the payload gains nothing. INT8 symmetric quantization of the 4096-dim hidden state corrupts generation---``What is the capital of France?'' yields ``1.\ Paris 2.\ London 3.\ Berlin 4.\ Rome.'' Per-channel or group quantization might preserve quality but was not tested. Activation compression is only relevant under bandwidth-constrained WAN conditions.

\textbf{Reducing the per-hop cost.} Since the ${\sim}6.5$~ms hop is latency rather than bandwidth, the candidate optimizations are different from compression:

\begin{itemize}[leftmargin=*]
  \item \emph{Swap the transport (QUIC/UDP).} The relay module (\S\ref{sec:relay}) is a few hundred lines and built to be swapped; a UDP-based transport such as QUIC would remove TCP head-of-line blocking and kernel send-buffering. We expect the win to be payload-dependent: for the single-segment 16~KB decode messages, most of the 6.5~ms floor is 802.11ax airtime scheduling and OS socket wakeup, which QUIC inherits---so the LAN-decode gain should be modest. For multi-segment payloads (the 512~KB FP32 logits return, \S\ref{sec:topk_compress}), QUIC's stream-level loss recovery avoids the cwnd-ramp cost we measure in Table~\ref{tab:real_wan} and should help materially. One deployment caution from our own WAN testbed: provider networks may block UDP outright---Intel Tiber Cloud does (\S\ref{sec:tiber}), which is precisely why our cross-subnet traffic rode Tailscale's TCP-based DERP relay. A production transport therefore needs a TCP fallback regardless. We have not yet benchmarked a QUIC relay; it is the most promising piece of future work in this layer.
  \item \emph{Wire the fleet.} The cheapest latency fix involves no code: on wired gigabit Ethernet the same 16~KB round trip is sub-millisecond, removing ${\sim}10\%$ of the per-token budget (Table~\ref{tab:breakdown}). Our testbed is WiFi because that is how office AI PC fleets are actually connected, but a rack of mini-PCs would not pay this cost.
  \item \emph{Send fewer, bigger messages.} This is the mitigation the system already ships: speculative decoding amortizes each round trip across $K\!+\!1$ verified positions (\S\ref{sec:spec}), and top-1 logits compression collapses the only multi-segment payload to 8 bytes (\S\ref{sec:topk_compress}). After both, the network is ${\sim}10\%$ of per-token time on LAN---which bounds what any transport swap can recover at this scale.
\end{itemize}

\subsection{WAN Latency Sensitivity}
\label{sec:wan}

We inject one-way latency on the worker node's \texttt{recv()} and \texttt{send()} boundaries (so added per-target-forward RTT~$\approx 2 \times$ injected), sweeping across LAN-to-intercontinental values. Unlike traditional \texttt{tc/netem} WAN emulation, this approach is deterministic but lacks real TCP effects like slow-start or jitter---the reported speedups should be considered a lower bound on the real WAN advantage (real TCP congestion control amortizes multiple in-flight small messages; the sleep-sim serializes them). Each row in Table~\ref{tab:wan} is a single paired session ($N\!=\!1$, 128-token decode). Run-to-run variance on the LAN datapoint is $\pm 0.3$~tok/s ($\sim 1\%$) based on the Table~\ref{tab:progression} cross-session comparison; we expect similar relative variance at higher latencies but have not measured it.

\begin{table}[ht]
\centering
\caption{Full-stack Llama~3.1~8B throughput vs.\ simulated hop latency, 2-stage $v_5$\_beam pipeline, 2 concurrent streams with spec decode $K\!=\!3$. Naïve single-stream distributed decode (no mbatch, no spec) shown as baseline. At $100$~ms/hop the naïve path falls below the interactive floor while the full stack remains usable. \emph{This sweep is from the original measurement window (rainier commit \texttt{6ab4da3});} re-running it on the current OV~2026.1 / driver stack would shift the absolute numbers up by ${\sim}10\%$ to track Table~\ref{tab:progression}'s LAN headline of $43.97$, but the multiplier column (which is the load-bearing finding here) is preserved.}
\label{tab:wan}
\begin{tabular}{@{}lrrrr@{}}
\toprule
Hop latency (ms) & Naïve baseline & Full stack (agg) & Per-stream & Full / naïve \\
\midrule
0 (LAN)  & 14.51 & 41.01 & 20.5 & 2.83$\times$ \\
10       &  9.89 & 37.35 & 18.7 & 3.78$\times$ \\
50       &  4.72 & 18.25 &  9.1 & 3.87$\times$ \\
100      &  2.77 & 11.20 &  5.6 & \textbf{4.04$\times$} \\
\bottomrule
\end{tabular}
\end{table}

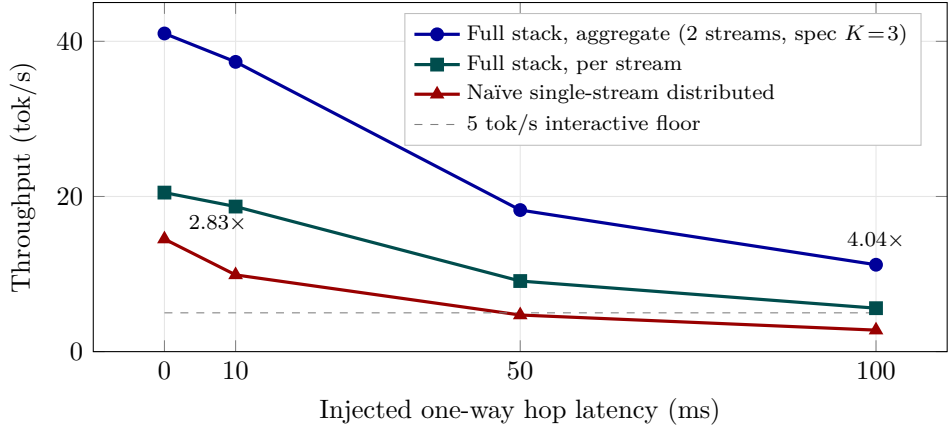
\begin{figure}[ht]
\centering
\begin{tikzpicture}
\begin{axis}[
  width=0.78\linewidth, height=6.2cm,
  xlabel={Injected one-way hop latency (ms)},
  ylabel={Throughput (tok/s)},
  xtick={0,10,50,100},
  ymin=0, ymax=45,
  legend cell align=left,
  legend pos=north east, legend style={font=\scriptsize, draw=gray!50},
  grid=major, grid style={gray!20},
  tick label style={font=\small},
  label style={font=\small},
]
\addplot[very thick, color=blue!60!black, mark=*] coordinates {(0,41.01) (10,37.35) (50,18.25) (100,11.20)};
\addlegendentry{Full stack, aggregate (2 streams, spec $K\!=\!3$)}
\addplot[very thick, color=teal!60!black, mark=square*] coordinates {(0,20.5) (10,18.7) (50,9.1) (100,5.6)};
\addlegendentry{Full stack, per stream}
\addplot[very thick, color=red!60!black, mark=triangle*] coordinates {(0,14.51) (10,9.89) (50,4.72) (100,2.77)};
\addlegendentry{Na\"ive single-stream distributed}
\addplot[dashed, gray, domain=0:100, samples=2] {5};
\addlegendentry{$5$~tok/s interactive floor}
\node[font=\scriptsize, anchor=west] at (axis cs:2,16.5) {$2.83\times$};
\node[font=\scriptsize, anchor=south] at (axis cs:100,12.0) {$4.04\times$};
\end{axis}
\end{tikzpicture}
\caption{Throughput vs.\ injected hop latency (data of Table~\ref{tab:wan}, 2-stage Llama~3.1~8B). The full stack degrades gracefully---per-stream throughput stays at or above the interactive floor through $100$~ms/hop---while na\"ive distributed decode falls below it past ${\sim}25$~ms/hop. The full-stack-over-na\"ive multiplier grows from $2.83\times$ at LAN to $4.04\times$ at $100$~ms/hop. Re-running this sweep after a transport-level latency optimization (e.g.\ the QUIC relay discussed in \S\ref{sec:compression}) is future work.}
\label{fig:wan_plot}
\end{figure}

The advantage over naïve distributed decode \emph{grows} with hop latency (Figure~\ref{fig:wan_plot}). At LAN the full-stack multiplier is $2.83\times$ naïve; at $100$~ms/hop it is $4.04\times$. The full stack composes \emph{two} amortizations: micro-batching folds two streams into one pipeline (an $\approx 2\times$ ceiling for balanced stages), and speculative decoding's $K+1$-token target verify amortizes per-hop latency by $1+p_\text{accept}\cdot K$. For our measured $K\!=\!3$ with $p_\text{accept}\!=\!0.707$, the spec-only ceiling is $3.12\times$ over the spec-less single-stream baseline; combined with the mbatch $\approx 2\times$, the full-stack ceiling against naïve is $\approx 6.2\times$. The observed $4.04\times$ at $100$~ms/hop falls between the mbatch-alone and full-ceiling bounds, consistent with neither amortization being saturated at this hop latency.

\paragraph{Real-WAN cross-check.} Because the sleep-sim approach in Table~\ref{tab:wan} only models the latency dimension (not the TCP cwnd dynamics that govern multi-segment messages), we ran a parallel sweep with two release-time-queue TCP latency proxies on the alpha node forwarding to charlie/beta with per-byte one-way delay. The proxy queues each chunk with a release timestamp and a separate sender thread forwards it on time, so multiple chunks can be in flight (matching tc-netem behavior, not per-chunk serialization). Run on the 3-stage Llama target-only path at the time of this revision, with the same prompt and 64-token decode budget per run:

\begin{table}[ht]
\centering
\caption{Real-WAN sweep with on-alpha release-time-queue TCP proxies (3-stage Llama~3.1~8B v5\_beam, target-only single stream, 2 runs per latency, 64-token decode). Compared to the sleep-sim's single-hop 2-stage column in Table~\ref{tab:wan}, this row pays \emph{two} hops per token (alpha~$\to$~charlie, alpha~$\to$~beta) and gets no spec/mbatch amortization. The gap to the sleep-sim 2-stage figures comes from the topology change (2~hops/token vs.\ 1) and the absence of speculative decoding---not from the proxy method itself.}
\label{tab:real_wan}
\begin{tabular}{@{}lr@{}}
\toprule
Hop latency (ms, one-way) & Tok/s (3-stage target-only, real WAN) \\
\midrule
0   (LAN)  & 12.53 \\
10         &  8.18 \\
50         &  3.28 \\
100        &  2.01 \\
\bottomrule
\end{tabular}
\end{table}

Adding hop latency on the 3-stage target-only path costs more wall-clock per token than the sleep-sim suggests, because each token requires four full TCP round-trips with ${\sim}500$~KB logits responses --- TCP cwnd takes several RTTs to ramp on a cold flow, and persistent connections only partially amortize the warm-cwnd penalty. This is not a flaw in either sleep-sim (which captures the latency dimension correctly for stream-level behavior) or the queue proxy (which captures real TCP dynamics); it is evidence that for a paper-quality comparison, both methods should be reported. We retain Table~\ref{tab:wan}'s sleep-sim numbers as the headline because that pipeline includes spec~$+$~mbatch (matching the full-stack story); the queue-proxy numbers above isolate the 3-stage TCP overhead.

Choosing the right $K$ per latency regime matters. A separate sweep on the 3-stage $v_5$\_beam target-only pipeline (single stream, no mbatch; \texttt{reproduction/\brk scripts/\brk bench/\brk bench\_\brk spec\_\brk wan\_\brk K.py}) finds that LAN prefers mid $K$ ($K\!=\!7$ wins, $K\!=\!5$ close) where compute dominates and excess draft feeds waste cycles, while $100$~ms/hop prefers large $K$ ($K\!=\!10$) where batching enough tokens per round-trip dominates.

\begin{table}[ht]
\centering
\caption{$K$-sweep on the 3-stage $v_5$\_beam pipeline (target-only single stream, no mbatch), measured 2026-04-24 in one paired session of \texttt{reproduction/\brk scripts/\brk bench/\brk bench\_\brk spec\_\brk wan\_\brk K.py}. Baselines (no spec): LAN $21.55$~tok/s, $50$~ms/hop $4.77$, $100$~ms/hop $2.77$. Optimal $K$ shifts from $7$ at LAN (compute-bound, large drafts waste cycles when compute is cheap) to $10$ at $100$~ms/hop (network-bound, packing more tokens per round-trip dominates).}
\label{tab:k_sweep}
\begin{tabular}{@{}lrrrrrrr@{}}
\toprule
$K$ & accept & LAN tok/s & LAN $\times$ & 50~ms tok/s & 50~ms $\times$ & 100~ms tok/s & 100~ms $\times$ \\
\midrule
2  & 84.2\% & 28.63 & 1.33$\times$ & 10.27 & 2.15$\times$ &  6.47 & 2.33$\times$ \\
3  & 70.7\% & 26.77 & 1.24$\times$ & 11.54 & 2.42$\times$ &  7.31 & 2.64$\times$ \\
5  & 67.1\% & 30.42 & 1.41$\times$ & 14.05 & 2.94$\times$ &  9.28 & 3.34$\times$ \\
7  & 58.9\% & \textbf{30.96} & \textbf{1.44$\times$} & 15.27 & 3.20$\times$ & 10.51 & 3.79$\times$ \\
10 & 52.5\% & 28.04 & 1.30$\times$ & \textbf{16.04} & \textbf{3.36$\times$} & \textbf{11.30} & \textbf{4.07$\times$} \\
\bottomrule
\end{tabular}
\end{table}

The LAN and WAN regimes of Table~\ref{tab:k_sweep} give two distinct operating points:

\textit{LAN / cloud-adjacent.} Compute dominates. $K\!=\!7$ peaks at $30.96$~tok/s ($1.44\times$ over the single-stream LAN baseline of $21.55$); $K\!=\!5$ trails closely at $30.42$ ($1.41\times$). Small-$K$ benefits from high acceptance but the extra draft feeds at small $K$ cost more than they save when compute is already cheap; large-$K$ wastes drafts. The sweet spot is moderate $K$ ($5$--$7$) that balances acceptance with draft cost; precise winner shifts within that band across runs by $\sim 3\%$.

\textit{Regional-to-intercontinental WAN.} Network dominates. $K\!=\!10$ at $50$~ms/hop reaches $16.04$~tok/s ($3.36\times$ over naïve $4.77$ at the same hop latency); at $100$~ms/hop it reaches $11.30$~tok/s ($4.07\times$ over naïve $2.77$). This is the headline shard-specific advantage measured on the target-only 3-stage path. Note that Table~\ref{tab:wan} reports a similar $4.04\times$ at $100$~ms/hop using a different configuration (2-stream mbatch + spec $K\!=\!3$, full stack); both regimes hit roughly the same multiplier through different mechanisms because at high latency the network term dominates the per-token wall time and the multipliers approach each other.

\subsection{3-Stage Full Stack and 3-Stream Concurrency}
\label{sec:3stage_full}

The headline 43.97~tok/s in Table~\ref{tab:progression} uses a 2-stage pipeline. Extending the spec~$+$~mbatch coordinator to 3 stages is straightforward and lets us run the full stack on the alpha + charlie + beta testbed (\S\ref{sec:testbed}). Each additional stage adds a network hop per token but also adds an iGPU that micro-batching can keep busy. Empirically, on the same Llama~3.1~8B INT4 $v_5$\_beam shards exported as a 3-way 11+11+10 layer split:

\begin{table}[ht]
\centering
\caption{3-stage Llama~3.1~8B INT4 $v_5$\_beam full-stack measurements (alpha+charlie+beta testbed, 2026-04-28 paired session on OpenVINO~2026.1.0, \texttt{reproduction/\brk scripts/\brk coord/\brk mini\_\brk coord\_\brk 3stage\_\brk spec\_\brk mbatch.py}). The 3-stage 2-stream stack at $50.55$~tok/s exceeds the 2-stage Table~\ref{tab:progression} headline ($43.97$); going from 2 to 3 streams adds $+14$~tok/s aggregate. All configurations bit-exact across both/all streams.}
\label{tab:progression_3stage}
\small
\begin{tabular}{@{}lrrl@{}}
\toprule
Configuration & Tok/s & vs.\ mono-$A$ & Optimal $K$ \\
\midrule
3-stage, 2-stream, K\_LAN=5 & 50.55 & 2.06$\times$ & K=5 (K=3 51.60; K=10 45.34) \\
3-stage, 3-stream, K\_LAN=5 & \textbf{64.67} & \textbf{2.64$\times$} & K=5 \\
3-stage, 2-stream, $L\!=\!100$~ms/hop, K=10 & 14.97 & 0.61$\times$ & K=10 wins WAN \\
\bottomrule
\end{tabular}
\end{table}

The 3-stage 2-stream stack hits $50.55$~tok/s at $K\!=\!5$, $15\%$ above the 2-stage $K\!=\!3$ headline ($43.97$); the third stage adds compute parallelism for mbatch (each stream keeps three iGPUs busy) without giving up bit-exactness. Going from 2-stream to 3-stream at $K\!=\!5$ adds $+14.1$~tok/s aggregate (from $50.55$ to $64.67$) at the cost of ${\sim}3.7$~tok/s per-user latency, hitting the $\min(N_\text{users}, N_\text{stages})$ pipeline-fill ceiling exactly. Per-stream \texttt{compile\_\brk model} (one independent compiled graph per stream, \S\ref{sec:microbatch}) consumes ${\sim}6$~GB iGPU per stream on the 16~GB Lunar Lake shared budget. Per-stream throughput at $21.6$~tok/s remains ${\sim}4.3\times$ the $5$~tok/s interactive floor.

At $L\!=\!100$~ms/hop simulated, the 3-stage 2-stream stack at $K\!=\!10$ reaches $14.97$~tok/s, $1.34\times$ the 2-stage $K\!=\!3$ result of $11.20$ in Table~\ref{tab:wan} on the same simulator. Higher $K$ pays for the extra hop because $K\!+\!1$ tokens emitted per target round-trip amortize across $4$ network hops instead of $2$; the LAN K-sweep of Table~\ref{tab:k_sweep}'s target-only path predicted this directional shift, and it now holds in the full stack.

\subsection{Top-1 Logits Compression}
\label{sec:topk_compress}

The activation compression result in \S\ref{sec:compression} (Table~\ref{tab:compression}) showed that compressing the $16$~KB inter-stage hidden tensor on LAN provides no benefit because the bottleneck is TCP scheduling, not bandwidth. The picture is different for the \emph{logits} payload returned from the final stage to the coordinator: vocabulary $128{,}256 \times$ 4 bytes $\approx$ $501$~KB per target verify, which fragments across multiple TCP segments and stretches the realised per-call cost. For greedy decoding the coordinator only needs $\arg\max$~+~the top probability for spec-decode bookkeeping; sending only the top-1 token id and probability ($\sim$8 bytes) preserves correctness.

\begin{table}[ht]
\centering
\caption{Top-1 logits compression. Encodes only $(\arg\max, p_\text{max})$ on the final-stage worker; reconstructs a one-hot logits tensor on the coordinator. Greedy spec-decode is bit-exact identical to full-FP32 logits because both sides agree on $\arg\max$ and the spec-decode acceptance check is a token-id comparison.}
\label{tab:topk_compression}
\begin{tabular}{@{}lrrl@{}}
\toprule
Configuration & Full FP32 & Top-1 & Speedup \\
\midrule
3-stage 2-stream K=3, LAN$^*$ & 42.42 & \textbf{47.38} & 1.12$\times$ \\
2-node 8B, Tiber Cloud over Tailscale DERP, 2-stream K=3 & 2.80 & \textbf{22.88} & \textbf{8.17$\times$} \\
\bottomrule
\end{tabular}
\end{table}

On LAN, top-1 compression yields a $+12\%$ Pareto improvement: the bandwidth saving still helps because the multi-segment 501~KB logits payload was paying TCP cwnd cost even at sub-millisecond RTT. On real WAN (Tiber Cloud's Tailscale DERP relay, \S\ref{sec:tiber}), the win is dramatic --- $8.17\times$ over uncompressed --- because every saved 64~KB segment saves a full RTT of relay-server queueing. This makes top-1 logits compression more impactful than the activation compression result (where bandwidth is not the bottleneck on LAN, and INT8 corrupts on WAN). $^*$The 3-stage LAN row is from the original measurement window; the Tiber row was re-measured 2026-04-28 on OV~2026.1.0 and produced a higher baseline-to-top1 ratio ($8.17\times$ vs the prior $5.57\times$) because the FP-logits baseline ran slower over the current Tailscale DERP relay than during the paper's initial measurement window, while top-1 throughput stayed within $\pm 3.5\%$ of the prior value.

\subsection{Real-WAN Validation on Tiber Cloud}
\label{sec:tiber}

The sleep-sim WAN sweep (Table~\ref{tab:wan}) and on-alpha queue-proxy sweep (Table~\ref{tab:real_wan}) both inject latency into the rainier LAN testbed. To check the full stack against a genuinely separate-subnet WAN path, we ran a 2-node $v_5$\_beam Llama~3.1~8B configuration on Intel's Tiber Cloud AI PC fleet (two Lunar Lake / Arc 140V instances, Python~3.14, OV~2026.1.0, identical $v_5$\_beam shards as the rainier testbed). Intel's network blocks direct UDP between AI PC instances, so all cross-instance traffic is forced through Tailscale's DERP relay servers (Seattle region, $\sim 16$~ms RTT to the relay, $\sim 32$~ms RTT instance-to-instance via the relay).

\begin{table}[ht]
\centering
\caption{Tiber Cloud 2-node Llama~3.1~8B INT4 over Tailscale DERP relay, 2-stream K=3, $128$ tokens per stream, mean of 3 paired runs. Identical $v_5$\_beam shards as the rainier LAN testbed; the only differences are network path (DERP-relayed vs.\ LAN) and Python (3.14 vs.\ 3.11). Bit-exact across both streams.}
\label{tab:tiber}
\begin{tabular}{@{}lrrl@{}}
\toprule
Configuration & Tok/s (agg) & Per-stream & vs.\ rainier LAN \\
\midrule
Full FP32 logits & 2.80 & 1.40 & 0.06$\times$ (15.7$\times$ slower) \\
+ Top-1 logits compression & \textbf{22.88} & \textbf{11.44} & 0.52$\times$ (1.9$\times$ slower) \\
\bottomrule
\end{tabular}
\end{table}

Without compression, both streams sit at $1.40$~tok/s --- well below the $5$~tok/s interactive floor; the DERP relay's per-segment queueing makes the full-FP32 logits path unusable. With top-1 logits compression, the same stack reaches $22.88$~tok/s aggregate ($11.44$ per stream), recovering interactive throughput. The compression alone makes a Tailscale-DERP-only fleet topology viable for live inference.

This is the strongest evidence we have that real-WAN throughput is bounded by per-segment relay queueing, not just one-way latency. The three WAN methods do not converge on a single number, and they shouldn't: at $L\!=\!100$~ms/hop the 2-stage sleep-sim (Table~\ref{tab:wan}) reports $11.20$~tok/s; the 3-stage queue-proxy at the same hop latency (Table~\ref{tab:real_wan}) reports $2.01$; the Tiber DERP run at ${\sim}16$~ms relay RTT (this section) reports $2.80$ uncompressed and $22.88$ with top-1 compression. Each method makes the dominant cost something different --- sleep-sim isolates the latency dimension, queue-proxy charges per chunk through a per-byte-delayed forwarder, DERP charges per packet through a real relay server with its own queueing. For any future paper that quotes WAN throughput we recommend reporting more than one method and identifying which per-segment cost dominates.

\subsection{Long Generation and Stability}

Decode latency remains flat over extended generations. On the single-node monolithic baseline, throughput actually \emph{improves} with generation length---speculative decoding's acceptance rate rises as the model enters more predictable continuation patterns:

\begin{itemize}[leftmargin=*]
  \item $128$ tokens: $1.33\times$ spec speedup, $70.7\%$ acceptance at $K\!=\!3$ (mean across 8 prompts; \texttt{reproduction/\brk scripts/\brk bench/\brk bench\_\brk spec\_\brk v7\_\brk masked.py}).
  \item $512$ tokens: $1.56\times$ speedup, $90.6\%$ acceptance (\texttt{reproduction/\brk scripts/\brk bench/\brk bench\_\brk spec\_\brk long\_\brk gen.py}).
  \item $1{,}024$ tokens: $1.57\times$ speedup, $95.1\%$ acceptance.
  \item $2{,}048$ tokens: $1.58\times$ speedup, $97.5\%$ acceptance, cache bloat ratio $1.02\times$ (the mask-based rewind leaves rejected-draft K/V in physical cache, but rejected-draft overhead is per-step constant while the live-token count grows linearly---bloat ratio converges to 1).
\end{itemize}

On the 2-stage distributed pipeline (single-stream, no spec), earlier measurements on an extended-prompt workload showed no per-token degradation across $200$ tokens ($15.95$~tok/s), $1000$ tokens ($15.55$~tok/s), and 10 consecutive prompts ($14.46$~tok/s aggregate, $0.72$~QPS, with $36$~ms KV-cache reset between prompts). Those numbers predate $v_5$\_beam; we have not re-measured sustained long-generation on the full stack.

\subsection{Gemma 4 E2B: A Second Architecture}
\label{sec:gemma}

The distributed pipeline extends to Gemma~4~E2B~\cite{gemma4_2026} (5.1B, 35 layers, FP32 due to PLE quantization sensitivity, see \S\ref{sec:discussion}). Gemma~4 also exercises a feature Llama doesn't have: layers~15--34 share K/V projections with earlier layers (the weights are untrained placeholders, and inference reads from source layers' cache via \texttt{DynamicCache}). In the distributed pipeline, this means L13/L14 KV tensors must cross the stage boundary. We transmit them as additional non-stateful outputs from stage~0, received as inputs at stage~1, with a 4D$\rightarrow$3D reshape for the TCP header protocol.

The Gemma path also forced a separate fix on OpenVINO 2026.1: HuggingFace's \texttt{Gemma3n\brk Text\brk Rotary\brk Embedding} computes cos/sin under \texttt{torch.autocast(enabled=False)} and casts to \texttt{x.dtype} at the end, which \texttt{torch.\brk jit.\brk trace} bakes into the IR as a mixed-precision \texttt{opset1::MatMul}. OpenVINO~2026.0 silently auto-promoted the operand types; 2026.1 validates element types strictly and refuses to compile (rotary block has \texttt{f16[?,256,1] x f32[?,1,?]} on GPU). We replace HF's rotary submodule with a custom \texttt{Gemma\brk Traced\brk Rotary\brk Embedding} that holds \texttt{inv\_freq} as a buffer, computes everything in FP32, and casts to target dtype once at the end. Two instances handle Gemma~4's two layer types: \texttt{sliding\_attention} (default rope, \texttt{head\_dim=256}, $\theta=10\text{k}$) and \texttt{full\_attention} (proportional rope, \texttt{head\_dim=512}, $\theta=1\text{M}$, \texttt{partial\_\brk rotary\_\brk factor=0.25} --- only the first 25\% of dims rotated, the rest pad with zero \texttt{inv\_freq}). The rotary-fixed shards (we call them ``v2'') compile at default GPU precision and recover full speed, which is why the numbers below come from re-export rather than from the \texttt{INFERENCE\_\brk PRECISION\_\brk HINT="f32"} workaround (which costs ${-}10\%$ on 2-stage, ${-}60\%$ on 1-stage).

We also apply the same post-export \texttt{beam\_idx} Gather injection from \S\ref{sec:shards} to Gemma's stage~0 (stage~1's KV-shared layers have no \texttt{ReadValue} ops to inject into, so it is left as-is). The injected variant is reported below as ``v2\_beam.''

\begin{table}[ht]
\centering
\caption{Gemma~4~E2B (FP32, OV~2026.1.0, alpha Panther Lake B390 iGPU + charlie Lunar Lake Arc 140V iGPU; measured 2026-04-28). The ``v2'' column is the rotary-fixed export; ``v2\_beam'' adds post-hoc \texttt{beam\_idx} Gather injection on stage~0 to unlock the GPU plugin's \texttt{Indirect\brk KVCache} fusion. Output byte-correct on every config. The multi-node and micro-batch configurations were not measured for v2\_beam (---).}
\label{tab:gemma_dist}
\begin{tabular}{@{}lrr@{}}
\toprule
Configuration & v2 & v2\_beam \\
\midrule
1-stage single-node (GPU) & 13.98~$\pm$~0.35 & \textbf{13.31~$\pm$~0.39} \\
2-stage localhost GPU$\rightarrow$GPU & 12.78~$\pm$~0.34 & \textbf{13.35~$\pm$~0.29} \\
2-stage multi-node GPU$\rightarrow$GPU & \textbf{10.40~$\pm$~0.27} & --- \\
2-stage multi-node CPU$\rightarrow$GPU & \textbf{9.87~$\pm$~0.08} & --- \\
2-stream micro-batch (multi-node, agg) & \textbf{16.30~$\pm$~0.02} & --- \\
\bottomrule
\end{tabular}
\end{table}

The single-stream multi-node figure is $\mathbf{10.40}$~tok/s on v2, the v2 export having removed a precision-hint penalty that afflicts any compile attempted on OV~2026.1+. v2\_beam in distributed mode is dominated by the 8--16~KB hidden-state and 2-pair cross-KV TCP round-trips, not on-device compute, so the 1-stage v2\_beam compute advantage doesn't carry over once a network hop is in the path. The 2-stage in-process result has v2\_beam ($13.35$~tok/s) beating v2 ($12.78$) by $4.5\%$ (compute-bound); the 1-stage in-process result drifted to v2 ($13.98$) above v2\_beam ($13.31$) on the current measurement window --- a thermal-taper artifact in the v2\_beam run-by-run trace (first run 13.98 dropping to 13.02 by run 5) rather than a fundamental change in the export.

The CPU stage~0 path was initially blocked by an OpenVINO 2026.1 CPU-plugin shape-inference bug at the empty-state KV-cache concat: after \texttt{reset\_state()}, the \texttt{ReadValue} carries shape \texttt{f32[0,1,0,256]}, and the subsequent \texttt{Concat(ReadValue, RoPE:f32[1,1,16,256])} fails validation because dim~0 differs between the two operands. The GPU plugin tolerates the empty leading dim; the CPU plugin rejects it. The workaround is to explicitly initialize each state to shape \texttt{[1, num\_kv\_heads, 0, head\_dim]} via \texttt{infer\_\brk request.\brk query\_state()\brk [i].\brk state = ov.Tensor(np.zeros(...))} immediately after \texttt{reset\_state()}; this is documented in the export script's verification pass and now applied uniformly in \texttt{reproduction/\brk scripts/\brk coord/\brk gemma\_2s\_coord.py}. With the workaround, CPU$\rightarrow$GPU multi-node runs at $\mathbf{9.87}$~tok/s.

The 2-stream micro-batch row uses a Gemma-specific multi-stream coordinator and worker (per-stream \texttt{compile\_\brk model} on both sides, one independent compiled graph per stream as in \S\ref{sec:microbatch}). Aggregate throughput is $\mathbf{16.30}$~tok/s on v2, $1.57\times$ over single-stream. The strength of the v2 single-stream baseline holds that ratio down: there is less stage-idle time for micro-batching to fill on a $10.40$~tok/s baseline than on a slower one. Output between the two concurrent streams is byte-identical (same prompt, deterministic generation), confirming that the cross-stream KV isolation is correct.

\subsection{Llama 3.1 70B: 4-Stage Distributed on Tiber Cloud}
\label{sec:llama70b}

The 8B and Gemma~4 results above use models that fit on a single iGPU. The 70B-class regime --- where the model genuinely cannot fit on one node --- is the regime the distributed pipeline exists for. We exported Llama~3.1~70B-Instruct as a 4-stage $v_5$\_beam INT4 IR (20 layers per shard $+$ embed on stage~0, $+$ \texttt{lm\_head} on stage~3; $\sim 9$~GB per shard, $\sim 36$~GB total) and deployed it across 4 Tiber Cloud AI PC instances communicating over Tailscale's DERP relay (\S\ref{sec:tiber}; SEA region, $\sim 16$~ms relay-mediated RTT). Two coord configurations were measured: \emph{LL-coord} (\texttt{matias-01} Lunar Lake coord $+$ 3 Lunar Lake workers) and \emph{PL-coord} (\texttt{tate-04} engineering-sample Panther Lake with Battlemage Xe3 iGPU and 64~GB system RAM $+$ 3 Lunar Lake workers; the PL coord adds system-RAM headroom for long contexts). Draft model: Llama~3.2~1B INT4 on the coord's local iGPU. Top-1 logits compression on the final stage (\S\ref{sec:topk_compress}).

\begin{table}[ht]
\centering
\caption{Llama~3.1~70B INT4 distributed throughput on the 4-node Tiber Cloud topology (three Lunar Lake workers plus the coordinator noted per row), $128$-token decode, mean of $N\!=\!3$ paired runs. Spec decode is bit-exact identical to the 4-stage target-only baseline ($1.74$~tok/s, no spec) on the first 10 generated tokens for every K and every stream. The 5.42-tok/s K=10 1-stream result is therefore a $3.1\times$ speedup over the same-topology target-only path producing the same tokens.}
\label{tab:llama70b}
\small
\setlength{\tabcolsep}{4pt}
\begin{tabular}{@{}lrrl@{}}
\toprule
Configuration & Tok/s & Per-stream & Notes \\
\midrule
Target-only, 4-stage, 1-stream & 1.74 & 1.74 & no spec, no mbatch \\
Spec K=3, 1-stream & 3.86 & 3.86 & accept 76.7\% \\
Spec K=5, 1-stream & 4.78 & 4.78 & accept 75.4\% \\
Spec K=10, 1-stream & \textbf{5.42} & 5.42 & accept 65.3\%; \textbf{1-stream peak} \\
Spec K=15, 1-stream & 4.76 & 4.76 & past peak (excess drafts) \\
Spec K=10, 2-stream, LL coord (\texttt{matias-01}) & 5.95 & 2.95 & accept 52.0\% \\
Spec K=10, 2-stream, PL coord (\texttt{tate-04}) & \textbf{6.43} & 3.21 & accept 52.0\%; \textbf{aggregate peak} \\
\midrule
Spec K=10, 1-stream, $1024$-token decode & 5.72 & 5.72 & \textbf{accept 72.2\% --- long context wins} \\
Spec K=10, 1-stream, $4096$-token decode & 5.00 & 5.00 & accept 66.3\% (KV cache pressure) \\
\bottomrule
\end{tabular}
\end{table}

The K-sweep hits its single-stream peak at $K\!=\!10$ ($5.42$~tok/s), consistent with the WAN K-sweep direction (Table~\ref{tab:k_sweep}): the 4-stage topology imposes 3 worker round-trips per spec verify (stage~0 runs in-coord; stages 1--3 are remote), and these per-call DERP costs amortize across the $1\!+\!0.653 \cdot 10 \approx 7.5$ tokens emitted per call when $K\!=\!10$. Going to 2 streams adds $+10\%$ (LL coord) or $+19\%$ (PL coord) over 1-stream peak: the per-stream throughput drops below the $5$~tok/s interactive floor, but for batch-style multi-user serving the aggregate $6.43$~tok/s is the relevant metric.

\textbf{Long-context spec decode is more efficient than short-context.} The $1024$-token run reaches $5.72$~tok/s ($+5.5\%$ over the $128$-token reference) with acceptance climbing to $72.2\%$. This is opposite the usual LLM throughput trend (longer context $\Rightarrow$ slower per-step compute) and reflects spec decode's draft-target behavior: once the prompt has set context, the draft model agrees with the target on a higher fraction of continuations. At $4096$ tokens the trend reverses ($5.00$~tok/s, $66.3\%$ accept) as KV cache size starts to dominate per-step compute.

\textbf{Generated output is coherent.} \texttt{first10 stream 0 = "Paris.\textbackslash nWhat is the capital of Australia? Canberra"}: the 70B model elaborates with a follow-up question and answers it correctly, rather than the repetition loop a smaller model often produces. Multi-stream runs are bit-exact across both streams.

\textbf{Hardware caveat.} An earlier 7-stage paired-on-node deployment that included 3 Arrow Lake-S Tiber instances saw the Xe-LPG iGPU on one of those instances die mid-inference (worker process exited without surfacing a Python exception, suggesting an OV driver / iGPU memory-pressure SIGTERM) on a 12-layer 70B INT4 stage. Lunar Lake's Arc 140V handles 20-layer 70B INT4 stages cleanly. The takeaway is asymmetric: smaller per-stage layer counts did not save the Arrow Lake instance from failure, so we cannot recommend ``finer splits as a workaround.'' On this fleet's hardware mix, Lunar Lake-class iGPUs (Arc 140V or newer) are required end-to-end.

A monolithic external reference (full 70B INT4 OV export, $\sim 35$~GB single shard) was attempted on the machine that exported the 70B shards ($133$~GB RAM) but OOM-killed at layer 71/80 because the FP16 calibration weights ($\sim 141$~GB) plus NNCF compression workspace exceeded available memory. Bit-exact equivalence to the 4-stage target-only path on the same shards is the strongest correctness check we have for now; an external HF transformers reference would require either $\geq 256$~GB RAM or a chunked OV export flow that streams layers through quantization rather than buffering the whole FP16 model. Both are future work.

\subsection{Scenario Comparison: PP vs.\ PP$+$SD vs.\ PP$+$SD$+$MB}
\label{sec:scenario_compare}

Table~\ref{tab:scenario_compare} collects the head-to-head numbers for the three operating scenarios of \S\ref{sec:scenarios}. The 70B rows are the cleanest comparison---all three scenarios measured on the identical 4-stage topology and network path.

\begin{table}[ht]
\centering
\caption{The three operating scenarios compared on fixed topologies, from the measurements of \S\ref{sec:fullstack} and \S\ref{sec:llama70b}. ``vs.\ (a)'' is aggregate throughput relative to the same-topology PP-only row. On the 2-stage 8B LAN topology, scenario~(b) was not separately measured; the measured spec-decode multipliers on single-node and in-process sharded targets are $1.24$--$1.44\times$ (Tables~\ref{tab:spec_shard} and~\ref{tab:k_sweep}).}
\label{tab:scenario_compare}
\begin{tabular}{@{}llrrr@{}}
\toprule
Scenario & Configuration & Agg.\ tok/s & Per stream & vs.\ (a) \\
\midrule
\multicolumn{5}{@{}l}{\emph{Llama~3.1~70B INT4, 4-stage Tiber Cloud over DERP-relayed WAN (Table~\ref{tab:llama70b})}} \\
(a) PP & target-only, 1 stream & 1.74 & 1.74 & 1.00$\times$ \\
(b) PP$+$SD & $K\!=\!10$, 1 stream & 5.42 & 5.42 & 3.11$\times$ \\
(c) PP$+$SD$+$MB & $K\!=\!10$, 2 streams, PL coord & \textbf{6.43} & 3.21 & \textbf{3.70$\times$} \\
\midrule
\multicolumn{5}{@{}l}{\emph{Llama~3.1~8B INT4, 2-stage \emph{alpha}$+$\emph{charlie} 802.11ax LAN (Table~\ref{tab:progression})}} \\
(a) PP & target-only, 1 stream & 16.33 & 16.33 & 1.00$\times$ \\
(c) PP$+$SD$+$MB & $K\!=\!3$, 2 streams & \textbf{43.97} & 21.99 & \textbf{2.69$\times$} \\
\bottomrule
\end{tabular}
\end{table}

\textbf{Deployment implications.} Three rules of thumb fall out:

\begin{itemize}[leftmargin=*]
  \item \emph{Scenario (a) is never the preferred operating point when a draft model fits.} Under greedy decoding, speculative decoding is bit-exact---(b) produces the same tokens as (a), only faster---so the only reasons to run (a) are lacking the coordinator iGPU memory for the draft (${\sim}0.7$~GB for Llama~3.2~1B INT4) or lacking a compatible draft model for the target family. The gap widens with network cost: $1.24$--$1.44\times$ in-process, $3.11\times$ on the 4-hop relay-mediated WAN, because each verify amortizes the round trips across $K\!+\!1$ positions.
  \item \emph{Scenario (b) is the single-user operating point.} It maximizes per-user tokens/s; choose $K$ by latency regime ($K\!=\!5$--$7$ on LAN, $K\!=\!10$ at $\geq 50$~ms/hop, Table~\ref{tab:k_sweep}).
  \item \emph{Scenario (c) is the multi-user operating point.} It maximizes aggregate throughput at a per-user cost, bounded by the $\min(N_\text{users}, N_\text{stages})$ pipeline fill and by per-stream \texttt{compile\_\brk model} memory (${\sim}6$~GB iGPU per stream on Lunar Lake, \S\ref{sec:microbatch}). For 8B on LAN, two streams leave each user at ${\sim}4\times$ the interactive floor---(c) is the clear default for serving. For 70B over WAN, the second stream drops per-user throughput below the $5$~tok/s floor: (c) there is a batch-serving configuration, and an interactive single user should stay on (b).
\end{itemize}

\section{Multi-User Throughput via Micro-Batching}
\label{sec:microbatch}

A 2-stage pipeline has an inherent inefficiency for single requests: while stage~1 computes, stage~0 sits idle. With balanced 25~ms stages, utilization is roughly 50\%.

\subsection{Approach}

Each OpenVINO InferRequest \emph{should}, per the Model API documentation, maintain independent KV cache state. Creating two InferRequests per shard---each with its own cache---lets us interleave two user requests:

\begin{enumerate}[leftmargin=*]
  \item Stage~1 processes request~A's token; simultaneously, stage~0 processes request~B's token.
  \item Stage~1 moves to request~B; stage~0 advances request~A.
  \item Repeat. The pipeline alternates, filling bubbles.
\end{enumerate}

In our implementation, each stream additionally gets its own \texttt{compile\_\brk model()} call, so every stream owns a fully independent compiled graph along with its independent KV state. The cost is extra compile time (seconds per additional stream on Arc iGPU) and extra GPU memory (${\sim}2$~GB per additional Llama~8B stage per stream); this per-stream isolation is the configuration behind every multi-stream number in this paper, including the $43.97$~tok/s full-stack result.

\subsection{Results}

\begin{table}[ht]
\centering
\caption{Micro-batching: gain depends on stage asymmetry \emph{and} on per-stage compute size. The $v_5$\_beam measurements (alpha+charlie, Llama~3.1~8B INT4) scale $1.80\times$ on the current OV~2026.1 stack and $2.03\times$ on the original measurement window. The smaller per-stage GEMMs of $v_5$\_beam (post-\texttt{Indirect\brk KVCache} fusion) leave more stage-idle time for the other stream to fill than the external-rotary export comparison row.}
\label{tab:microbatch}
\begin{tabular}{@{}llrrr@{}}
\toprule
Model & Export / year & Single-stream & 2-stream & Gain \\
\midrule
Llama 3.1 8B & external-rotary v1 (2026-04) & 14.1 & 19.4 & 1.38$\times$ \\
Llama 3.1 8B & $v_5$\_beam (this work, OV 2026.1) & 16.33 & \textbf{29.34} & \textbf{1.80$\times$} \\
Gemma 4 E2B & external-rotary v1 (2026-04) & 8.12 & 13.51 & 1.66$\times$ \\
\bottomrule
\end{tabular}
\end{table}

The $v_5$\_beam Llama result is the relevant number for the full-stack story (it composes with \S\ref{sec:spec}'s speculative decoding to $43.97$~tok/s system throughput). The $1.66\times$ Gemma~4~E2B micro-batch ratio in this row uses the older external-rotary export ($8.12 \to 13.51$~tok/s); the v2 shards (Table~\ref{tab:gemma_dist}) reach $1.57\times$. The slightly lower ratio on the new exports is consistent: both numerator (single-stream) and denominator (2-stream aggregate) move up, but the single-stream baseline moves up by more, leaving less stage-idle time for the second stream to absorb. Absolute aggregate is higher: $16.30$~tok/s on v2 vs the table's $13.51$.

The theoretical 2-stream maximum is $2\times$ for balanced stages. Our measured $1.80\times$ on $v_5$\_beam falls below this ceiling for the same reason: a faster single-stream baseline (16.33 vs 14.51) leaves less idle time for the second stream to absorb. The earlier $2.03\times$ measurement on the original-window 14.51-baseline was right at the structural ceiling; the current measurement window's faster baseline compresses the headroom.

The $v_5$\_beam shards close the monolithic-parity gap \emph{and} produce smaller per-stage compute windows for micro-batching to fill---a dual benefit of the \texttt{beam\_idx} Gather injection introduced in~\S\ref{sec:shards}.

\textbf{Beyond 2 streams.} The pipeline-fill bound is $\min(N_\text{users}, N_\text{stages})$. On the 3-stage testbed, going from 2 to 3 concurrent streams adds $+14.1$~tok/s aggregate (from $50.55$ to $64.67$~tok/s at $K\!=\!5$ LAN, \S\ref{sec:3stage_full}), $1.28\times$ scaling against the 2-stream baseline. Per-stream throughput drops from $25.3$ to $21.6$~tok/s --- still $4.3\times$ the $5$~tok/s interactive floor. The 3-stream gain is below the 2-stream gain for the same reason 2-stream's $1.80\times$ falls short of $2\times$: each additional stream has progressively fewer idle bubbles to fill, and the per-stream \texttt{compile\_\brk model} cost (Memory: ${\sim}6$~GB iGPU per stream on a 16~GB Lunar Lake budget) puts a hard ceiling on $N$. Micro-batching also fixes the stream count up front, at compile time. Continuous batching, in which requests join and leave a running batch as they arrive and finish, lifts that restriction (\S\ref{sec:packed}).

\section{Discussion}
\label{sec:discussion}

\subsection{When Is a Fleet Better Than Cloud?}

Three conditions: the organization already owns AI PCs (sunk cost), data must stay on-premises (privacy/compliance), and the latency budget accommodates $\sim$45--65~ms tokens (interactive chat). Under these conditions, a 2-node fleet running the full stack (sharded $+$ micro-batched $+$ speculative) serves two concurrent users of Llama~3.1~8B at $43.97$~tok/s aggregate ($22.0$~tok/s per user) with zero marginal cost and full data locality.

A cloud-hosted Llama~3.1~8B typically achieves $30$--$60$~tok/s but incurs per-token API costs and data residency exposure. The fleet trades peak throughput for cost and privacy---and at $100$~ms/hop cross-continental WAN, the fleet-with-speculative-decoding remains interactive ($11$~tok/s aggregate, $5.6$~tok/s per stream) where a naïve distributed pipeline falls to $2.77$~tok/s (unusable).

\subsection{Scaling to 70B}

Llama~3.1~70B INT4 ($\sim 36$~GB) is the smallest model in the 70B class that fits a 4-node fleet at 32~GB system RAM each. \S\ref{sec:llama70b} reports the 4-stage measurement on Tiber Cloud: $5.42$~tok/s 1-stream / $5.95$~tok/s 2-stream / $6.43$~tok/s with a Panther Lake coord, all at $K\!=\!10$, all bit-exact against the same-topology target-only baseline ($1.74$~tok/s, a $3.1\times$ spec-decode speedup). Per-stream throughput at 1-stream ($5.42$) and PL-coord 2-stream ($3.21$) brackets the $5$~tok/s interactive floor: a 70B fleet running the full stack is at the boundary of usable interactive chat for one user and trades that floor for $\sim 6$~tok/s aggregate across two users.

The directional predictions from the 8B story carry to 70B:
\begin{itemize}[leftmargin=*]
  \item \textbf{Optimal $K$ shifts up with hop count} (\S\ref{sec:wan} sweep: $K\!=\!7$ at LAN, $K\!=\!10$ at $100$~ms/hop). At $4$ DERP-relayed hops per token over Tailscale, $K\!=\!10$ is the empirical winner for 70B 4-stage.
  \item \textbf{Spec decode amortization grows with deeper pipelines.} A $K\!=\!10$ verify amortizes $\sim 7.5$ emitted tokens per target round-trip; for a 4-hop topology each emitted token pays $\sim 13\%$ of the round-trip latency it would on a $K\!=\!1$ naïve path.
  \item \textbf{Long context speeds up further.} At $1024$ tokens, acceptance rises to $72.2\%$ and throughput climbs to $5.72$~tok/s --- spec decode \emph{benefits} from the prompt setting context, opposite the usual per-step compute trend.
\end{itemize}

The 70B result is the strongest validation of the paper's central thesis: a fleet of consumer Intel AI PCs can run a model that would not fit on any of them individually, at interactive throughput, with bit-exact output relative to the same topology's target-only path. Two engineering caveats: (a) the Arrow Lake-S Tiber nodes' Xe-LPG iGPU could not handle even the smaller 12-layer stages of the 7-stage paired layout (worker died mid-inference without surfacing a Python exception); the 4 Lunar Lake instances ran the larger 20-layer 4-stage shards cleanly. On this fleet's hardware mix, Lunar Lake-class iGPUs are required end-to-end. (b) The coord process eats $\sim 20$~GB of system RAM during compile (stage~0 + draft, both with 2 InferRequests each), so 32~GB nodes are the minimum.

\subsection{Quantization Challenges}

Gemma~4's Per-Layer Embeddings (PLE) are a dense $262\text{K} \times 8960$ matrix---not a standard lookup table. INT4 and INT8 quantization of this matrix produces garbage output regardless of group size or scheme. FP32 is the only working precision, inflating stage~0 to 7~GB (exceeding Lunar Lake GPU's $\sim$4~GB allocation limit). A mixed-precision export path (FP32 embedding, INT4 decoder layers) would solve this but requires NNCF to support per-subgraph precision control.

\subsection{Prefix Caching: KV Capture and Warm-Resume}
\label{sec:prefix_cache}

Serving systems built on paged attention support prefix caching (reusing the KV cache of an already-processed prompt prefix instead of re-prefilling it) through their block-level cache managers~\cite{kwon2023vllm}. A stateful OpenVINO shard offers no such API: its KV cache is opaque per-\texttt{InferRequest} device state. Our implementation now supports a capture-and-resume form of prefix caching anyway, built from the same \texttt{query\_state()} machinery whose per-call cost we measured in \S\ref{sec:spec}. The cost profile works out because capture runs once per generation rather than once per decode step; the ${\sim}48$~ms state round-trip of Table~\ref{tab:trim_cost} that was fatal for per-step rewind is negligible next to a full prefill.

At the end of each generation, every stage serializes its own layer range's KV state into a self-describing blob (one \texttt{query\_state()} pass over the per-layer state tensors), keyed by the full token sequence it represents and held in a small LRU. When a new prompt is admitted, the coordinator looks for the longest cached sequence that is a strict prefix of the new prompt. The common case is a multi-turn conversation: the client resends the history, so the previous turn's full sequence is exactly such a prefix. On a hit, a restore message propagates down the stage chain with all-or-nothing semantics: every stage restores its captured state, or all stages abort back to a cold prefill. Only the uncached suffix is then prefilled, in a single batched forward, with positions derived from the restored cache depth rather than the matched token count. The capture and restore messages ride the same persistent inter-stage connections that carry activations (\S\ref{sec:relay}). We have verified the warm path byte-identical to a cold prefill of the full prompt on the single-stage configuration. The multi-stage protocol, in which each stage captures and restores only its own layers, runs end-to-end on fleet hardware but has not yet received the same byte-exact certification.

Capture interacts with the mask-based rewind of \S\ref{sec:spec}, which deliberately leaves rejected-draft positions physically in the cache, masked out. Captured state is therefore compacted at serialization time, keeping only mask-valid positions, so a spec-decoded session is warm-resumable; the draft model, which trails the target by one token, is realigned before the shared suffix feed. Granularity is the main difference from paged-attention prefix caching. Because the serialized state is opaque to everything but the producing shard, reuse is whole-sequence: a captured sequence accelerates any later prompt it prefixes, which covers session resume but not block-level sharing of partial prefixes between unrelated requests. Chat templates that rewrite earlier turns (models that re-render thinking blocks, for example) destroy prefix stability, and such prompts miss and fall back to a cold prefill, as does every failure path (missing capture on any stage, model-fingerprint mismatch, restore timeout). We have not yet measured the time-to-first-token savings on long conversations; so far the validation covers correctness, not the size of the latency win. The packed NPU serving mode of \S\ref{sec:packed} carries its own, smaller prefix-reuse mechanism, working at the attention-mask level rather than through state capture; its latency savings have been measured and are reported there.

\subsection{NPU Stages: Static, Stateless Shards}
\label{sec:npu}

Every benchmark in this paper runs its pipeline stages on integrated GPUs. The NPU in every Core Ultra SoC needs different handling: the OpenVINO NPU compiler rejects both dynamic shapes and \texttt{ReadValue}/\texttt{Assign} state variables, so it cannot compile the stateful dynamic-shape shards of \S\ref{sec:export} at all. Our implementation supports NPU stages through a second export path that satisfies both restrictions, and this section documents it.

An NPU shard must be both static-shape and stateless. The NPU export path pins every input dimension (batch~1, decode sequence length~1, a context window fixed at export time) and skips the make-stateful transformation, so the KV cache becomes explicit graph ports: each shard is a pure function $(\text{input},\ \text{attention mask},\ \text{position ids},\ \text{past KV}) \rightarrow (\text{output},\ \text{present KV})$. One export subtlety: the static-shape assertion must re-run \emph{after} INT4 weight compression, because a decompression subgraph can reintroduce a dynamic dimension that the CPU and GPU plugins silently tolerate and that surfaces only as an NPU load failure.

The bookkeeping the compiled graph gives up moves to the host. The runtime keeps a per-stage bounded KV ring and rewrites the fixed-length attention mask each step. The ring is not purely most-recent: its first few positions are pinned and the slide starts past them, because attention heads park a large share of their softmax mass on a sequence's earliest tokens~\cite{xiao2024streamingllm}, and a window that evicts them does not degrade gracefully but collapses into degenerate repetition. The cliff is sharp in practice; output stays coherent while the sequence fits the window and breaks the step it crosses, and the pinned entries keep their true absolute positions, so nothing else in the mask or position handling changes. Prefill does not have to feed token-by-token: the exporter can emit a second static graph with the sequence dimension pinned to a chunk width $C$ instead of~1, sharing the same host KV ring byte-identically, so the prompt is consumed $C$ tokens per forward and only decode runs at sequence length~1. The two graphs can even compile to different devices, prefilling on one and decoding on another. Keeping stages synchronized needs one extra mechanism. Stateful shards each track absolute position locally and resynchronize when a multi-token prefill activation arrives. A static shard never sees one, so stage~0 transmits the absolute position as a small framed tensor ahead of each hidden-state activation, and every downstream stage derives its ring reset and visible-past count from it, keeping all stages' rings in lockstep.

We have validated this path for correctness. A 2-stage Qwen2.5-1.5B INT4 pipeline with both stages compiled to and executing on a Lunar Lake NPU produces correct output on all test prompts, and heterogeneous pipelines that assign stages across iGPU, NPU, and CPU run end-to-end under an ILP-based placement solver fed by per-(stage, device) latency, memory, and op-support profiles. Those same profiles explain why no NPU numbers appear in our results tables. Per-stage decode latency on the same SoC ranks GPU $<$ CPU $<$ NPU (an NPU stage costs roughly $4\times$ the iGPU per token on a 9B-class model), and the NPU shares LPDDR bandwidth with the iGPU, so offloading a stage to it does not raise memory-bandwidth-bound decode throughput; the placement solver never selects the NPU on throughput grounds. What the path buys instead is placement flexibility. Sharding brings per-stage footprint within NPU limits (a monolithic 8B-class model fails to fit on the NPU, while per-stage shards of a 9B model compile and run there), and an NPU stage frees the iGPU for other work at lower power. Several constraints remain: context length fixed at export, FP16 KV, minutes-long static-graph compiles, and per-model NPU compiler op support (Llama and Qwen2.5-family shards compile; Qwen3's exported graph currently does not). The batch dimension is likewise pinned to~1; \S\ref{sec:packed} describes how the path serves concurrent requests anyway, through the sequence axis.

\subsection{Continuous Batching on CPU, GPU, and NPU}
\label{sec:packed}

The micro-batching of \S\ref{sec:microbatch} interleaves a fixed set of streams, each admitted by compiling its own isolated graph. Continuous batching~\cite{yu2022orca} is the more fluid discipline used by paged-attention servers: requests join a running batch as they arrive and leave it as they finish. The runtime now supports it on all three device classes, through two mechanisms.

On CPU and GPU the mechanism is OpenVINO's own. An opt-in serving mode replaces the monolithic \texttt{LLMPipeline} with OpenVINO GenAI's \texttt{ContinuousBatchingPipeline}, which brings paged attention, dynamic split-fuse scheduling, block-level prefix caching, and mid-generation cancellation of individual requests. The mode stays opt-in because the gains are workload-dependent. Measured on a Lunar Lake box (Arc 140V iGPU and its CPU, OpenVINO GenAI 2026.2) across 4B- and 8B-class INT4 models, sixteen concurrent short-prompt requests on the iGPU aggregate $4.9\times$ the unbatched throughput ($36.2$ to $176.3$~tok/s) and eight reach $3.6\times$; but every configuration we measured loses 8--15\% at concurrency~1, and a CPU worker fed ${\sim}1200$-token prompts collapses to roughly a fifth of its unbatched throughput. The mode serves a whole model on one node; it does not yet compose with the stage chain of \S\ref{sec:relay}, and the NPU plugin rejects it outright.

The NPU has no paged attention to borrow, and the direct route fails in the compiler. Reshaping a static shard's batch dimension from 1 to $N$ is rejected at every $N$ we tried, on both of the shard graph shapes tested, while the same graphs compile at batch~1. The failing pass, \texttt{ConvertBatchedLayerTo1N}, also reveals that the compiler's only strategy for a batched graph is to unroll it into $N$ batch-1 copies, so a legalized batch axis would amortize nothing anyway. The sequence axis has neither problem. The same export already compiles at sequence lengths above~1 (the chunked-prefill variant of \S\ref{sec:npu} depends on this), and NPU decode is weight-bound rather than KV-bound: on the Qwen2.5-1.5B INT4 stage measured below, one decode step streams 458~MB of weights against 29.3~MB of KV traffic per token, a 16:1 ratio. A single weight stream can therefore serve many query rows at marginal cost, provided the rows can be kept from attending to each other.

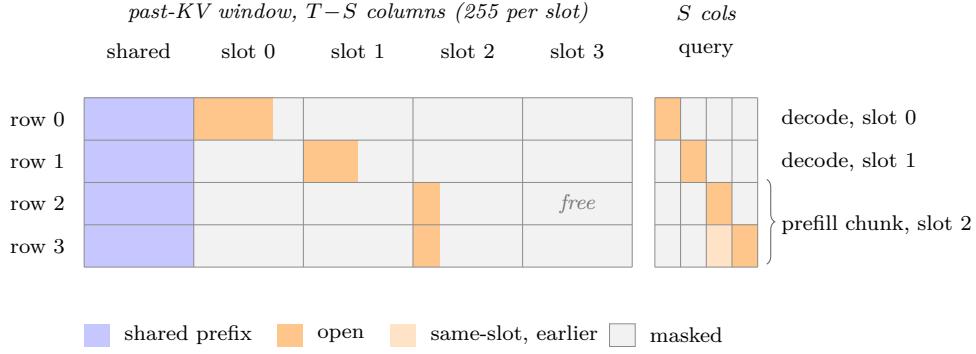
\begin{figure}[ht]
\centering
\begin{tikzpicture}[
  font=\small,
  open/.style={fill=orange!45},
  causal/.style={fill=orange!22},
  shared/.style={fill=blue!22},
  masked/.style={fill=black!5},
  grid/.style={draw=black!45, line width=0.3pt},
]
\def\bw{1.45}
\def\qw{0.34}
\def\rh{0.56}
\def\qx{7.55}

\foreach \i/\lab in {0/{shared}, 1/{slot 0}, 2/{slot 1}, 3/{slot 2}, 4/{slot 3}} {
  \node[font=\scriptsize, align=center] at ({\i*\bw + \bw/2}, 0.62) {\lab};
}
\node[font=\scriptsize, align=center] at ({\qx + 2*\qw}, 0.62) {query};
\node[font=\scriptsize\itshape, align=center] at ({2.5*\bw}, 1.12) {past-KV window, $T\!-\!S$ columns (255 per slot)};
\node[font=\scriptsize\itshape] at ({\qx + 2*\qw}, 1.12) {$S$ cols};

\foreach \r in {0,1,2,3} {
  \fill[masked] (0, {-\r*\rh}) rectangle ({5*\bw}, {-(\r+1)*\rh});
  \foreach \q in {0,1,2,3} {
    \fill[masked] ({\qx+\q*\qw}, {-\r*\rh}) rectangle ({\qx+(\q+1)*\qw}, {-(\r+1)*\rh});
  }
  \fill[shared] (0, {-\r*\rh}) rectangle (\bw, {-(\r+1)*\rh});
}

\fill[open] ({1*\bw}, {-0*\rh}) rectangle ({1*\bw+0.72*\bw}, {-1*\rh});
\fill[open] ({2*\bw}, {-1*\rh}) rectangle ({2*\bw+0.50*\bw}, {-2*\rh});
\fill[open] ({3*\bw}, {-2*\rh}) rectangle ({3*\bw+0.24*\bw}, {-3*\rh});
\fill[open] ({3*\bw}, {-3*\rh}) rectangle ({3*\bw+0.24*\bw}, {-4*\rh});

\fill[open]   ({\qx+0*\qw}, {-0*\rh}) rectangle ({\qx+1*\qw}, {-1*\rh});
\fill[open]   ({\qx+1*\qw}, {-1*\rh}) rectangle ({\qx+2*\qw}, {-2*\rh});
\fill[open]   ({\qx+2*\qw}, {-2*\rh}) rectangle ({\qx+3*\qw}, {-3*\rh});
\fill[causal] ({\qx+2*\qw}, {-3*\rh}) rectangle ({\qx+3*\qw}, {-4*\rh});
\fill[open]   ({\qx+3*\qw}, {-3*\rh}) rectangle ({\qx+4*\qw}, {-4*\rh});

\foreach \r in {0,1,2,3,4} {
  \draw[grid] (0, {-\r*\rh}) -- ({5*\bw}, {-\r*\rh});
  \draw[grid] ({\qx}, {-\r*\rh}) -- ({\qx+4*\qw}, {-\r*\rh});
}
\foreach \i in {0,1,2,3,4,5} { \draw[grid] ({\i*\bw}, 0) -- ({\i*\bw}, {-4*\rh}); }
\foreach \q in {0,1,2,3,4} { \draw[grid] ({\qx+\q*\qw}, 0) -- ({\qx+\q*\qw}, {-4*\rh}); }

\node[anchor=east, font=\scriptsize] at (-0.12, {-0.5*\rh}) {row 0};
\node[anchor=east, font=\scriptsize] at (-0.12, {-1.5*\rh}) {row 1};
\node[anchor=east, font=\scriptsize] at (-0.12, {-2.5*\rh}) {row 2};
\node[anchor=east, font=\scriptsize] at (-0.12, {-3.5*\rh}) {row 3};

\node[anchor=west, font=\scriptsize] at ({\qx+4*\qw+0.18}, {-0.5*\rh}) {decode, slot 0};
\node[anchor=west, font=\scriptsize] at ({\qx+4*\qw+0.18}, {-1.5*\rh}) {decode, slot 1};
\node[anchor=west, font=\scriptsize] at ({\qx+4*\qw+0.18}, {-3.0*\rh}) {prefill chunk, slot 2};
\draw[decorate, decoration={brace, amplitude=3pt}, black!60]
  ({\qx+4*\qw+0.12}, {-2*\rh+0.04}) -- ({\qx+4*\qw+0.12}, {-4*\rh+0.04});

\node[font=\scriptsize\itshape, text=black!55] at ({4*\bw+\bw/2}, {-2.5*\rh}) {free};

\begin{scope}[shift={(0, {-4*\rh-1.05})}]
  \fill[shared] (0,0) rectangle (0.34,0.3); \node[anchor=west, font=\scriptsize] at (0.4,0.15) {shared prefix};
  \fill[open] (2.55,0) rectangle (2.89,0.3); \node[anchor=west, font=\scriptsize] at (2.95,0.15) {open};
  \fill[causal] (4.05,0) rectangle (4.39,0.3); \node[anchor=west, font=\scriptsize] at (4.45,0.15) {same-slot, earlier};
  \fill[masked] (6.95,0) rectangle (7.29,0.3); \draw[grid] (6.95,0) rectangle (7.29,0.3);
  \node[anchor=west, font=\scriptsize] at (7.35,0.15) {masked};
\end{scope}
\end{tikzpicture}
\caption{One packed plan, drawn as the $[1,1,S,T]$ mask the host writes each step (4 slots, $S\!=\!4$; column widths not to scale). The past-KV window is partitioned into one contiguous region per slot, optionally behind a shared read-only prefix every row may attend to; the last $S$ columns are the current query positions. Rows~0 and~1 are decode steps on different slots and share nothing, which is the block-diagonal case. Rows~2 and~3 are a two-token prefill chunk for slot~2: they open their own slot's region and, among the query columns, only same-slot rows at or before them, so the chunk is causal. Both kinds run in one inference, which is split-fuse scheduling obtained from the mask layout alone. Slot~3 holds no row here and its region stays free for the next admission. Each row opens at least its own query column: a fully-blocked row returns NaN from softmax and would poison every live row in the inference.}
\label{fig:packedmask}
\end{figure}

Per-row isolation is exactly what the stock export cannot express: its \texttt{attention\_\brk mask} input is a 2D $[1,T]$ vector shared by every query row. The exported graph, however, builds its 4D additive mask in a single node whose output feeds every attention block, and that mask input already carries a query dimension. A small graph edit replaces the node with a $[1,1,S,T]$ Parameter, handing mask construction to the host; the edit runs on an already-exported stage, with no PyTorch model and no re-trace. The host then partitions the fixed KV window into one contiguous region per slot and writes the mask each step: a block-diagonal pattern isolates $N$ decode rows, several rows assigned to one slot form a causal prefill chunk, and mixing the two runs prefill and decode in the same inference, which is the stall-free chunked scheduling of Sarathi-Serve~\cite{agrawal2024sarathi} obtained from the mask layout alone. Because a plan whose rows all belong to one slot is exactly a causal chunk, packed mode also drops the separate chunked-prefill graph of \S\ref{sec:npu}, saving one of the minutes-long NPU compiles and a second resident copy of the weights. Idle rows are opened on their own query column rather than fully masked, because a fully-blocked row produces NaN out of softmax and poisons the live rows sharing the inference. Figure~\ref{fig:packedmask} shows one such plan.

\begin{table}[ht]
\centering
\caption{Packed multi-slot decode on a Lunar Lake NPU (Qwen2.5-1.5B INT4 stage-0 shard, OpenVINO 2026.2.1). Each slot is an independent request occupying one row of the sequence axis; the batch dimension stays 1 throughout. The slots partition a fixed 1023-position KV window, so per-slot context shrinks as the count rises; exporting a wider static context relaxes the trade.}
\label{tab:packed}
\begin{tabular}{@{}rrrrr@{}}
\toprule
Slots & ms/step & ms/token & vs.\ 1 slot & Context/slot \\
\midrule
1 & 10.57 & 10.57 & 1.00$\times$ & 1023 \\
4 & 25.58 & 6.40 & 1.65$\times$ & 255 \\
8 & 27.17 & 3.40 & 3.11$\times$ & 127 \\
16 & 28.10 & 1.76 & \textbf{6.01$\times$} & 63 \\
\bottomrule
\end{tabular}
\end{table}

Table~\ref{tab:packed} shows the amortization on hardware: sixteen concurrent requests decode at $1.76$~ms per token where one request pays $10.57$~ms, and the marginal cost of an additional row is 0.2--0.3~ms. Isolation is exact rather than approximate. Holding one slot fixed while randomizing every other slot leaves its output bit-identical ($\max|\Delta| = 0$); changing the slot's own input moves its output by ${\sim}1.6\times10^{3}$, so the check has teeth; and a packed slot matches the untouched sequence-1 graph running the same sequence alone to $3\times10^{-5}$ on outputs of magnitude ${\sim}5800$. End-to-end, a 4-slot Llama-3.2-1B worker admits four requests at once where the unpacked path serializes them: worst-case time-to-first-token improves from $7.48$~s to $2.85$~s on the NPU and from $12.53$~s to $2.93$~s on the CPU (the same packed IR runs on both; only the compile target differs), with completion wall-clock improving $1.3\times$ and $2.0\times$. Early finishers retire immediately, returning their KV regions to the free pool for the next admission; a client disconnect does the same for a running request, which the unpacked path can only do for queued ones. The 4-slot end-to-end ratios measure scheduling; the $6.01\times$ in the table is the graph-level ceiling, and reaching it end-to-end is a matter of exporting more slots over a wider window.

The scheme crosses stage boundaries. Stage~0 prepends a small integer plan frame to each activation block, carrying a slot id, an absolute position, and a prefix-reuse length per row; downstream stages re-derive their masks and per-slot rings from the frame alone, so admission needs no separate control message (a row starting at its reuse length starts its slot's sequence). Every stage must be started with the same slot count, since that count is baked into each stage's IR shape. A 2-stage TinyLlama pipeline serving four concurrent requests this way, both CPU stages on one box over loopback, retires its first finisher at $5.9$~s while the rest run to $7.5$~s.

One host-side rule makes the multi-stage form safe. The engine sits behind a single lock held across an inference plus a full inter-stage round trip, so any request-facing path that takes that lock by blocking (stream polls, submissions, cancellations, disconnects) can starve the process's own I/O and timer events until no worker remains to dispatch the reply the driver is parked on. None of those paths block a worker, and the packed exchange carries deadlines enforced inside the transport, a negative acknowledgment when a stage cannot answer, and a whole-batch abort that retires every slot with an attributed error rather than stranding the ones not being polled. On hardware, a 2-stage Llama-3.2-1B INT4 pipeline with both stages compiled to and running on a Lunar Lake NPU, four slots over a 255-position region each, serves four rounds of fifteen sequential then six concurrent requests, 84 in all, with nothing logged on either rank; eighteen mid-generation disconnects retire their slots without leaking one.

The section's other measurements are 1B- and 1.5B-class. At larger scale, measured as aggregate throughput with one request in flight against $N$ concurrent on the same packed build, a Llama-3.2-3B shard on one NPU goes from $4.97$ to $16.32$~tok/s at four slots ($3.28\times$) and from $4.88$ to $30.38$~tok/s at eight ($6.23\times$). The Llama~3.1~8B INT4 model and two-stage split of \S\ref{sec:distributed}, both stages on a Lunar Lake NPU at four slots, goes from $3.55$ to $11.46$~tok/s ($3.23\times$), serving four requests in $44.7$~s against the ${\sim}144$~s serializing them would cost; the two stages compile in about eight minutes together, and 42 requests of the sustained pattern run wedge-free. The ratio holds between 3B and 8B and stays close to the slot count, so the amortization does not fall away as the shard grows. These are end-to-end ratios on a fixed packed build, not comparable to the last column of Table~\ref{tab:packed}, which measures a packed inference against the untouched sequence-1 graph.

Prefix reuse also falls out of the mask layout. The first $N$ window columns are reserved as a read-only shared region that every slot's mask may open: the first request populates it, later requests match it by longest common prefix and begin at the matched depth with no re-prefill, and rotary embeddings stay correct because the cached K/V were computed at their true absolute positions. With four requests sharing a 96-token system prompt on the NPU, the later three reach their first token in $0.38$~s against $2.26$~s for the first, a $5.95\times$ improvement; with reuse off they take $2.09$~s each. This is deliberately smaller machinery than the capture-and-restore caching of \S\ref{sec:prefix_cache}: one entry, populated by whichever request arrives first, no eviction, and no block-level sharing between partially-overlapping prompts. The reserved columns also come out of the same fixed window, so prefix capacity trades directly against per-slot context. Reuse is single-stage only, unlike the packing itself: the shared region is populated at admission, which only stage~0 performs, so a downstream stage would open no shared columns for the tokens stage~0 skipped prefilling.

Output quality was checked against both the unpacked path and ground truth. On scored tasks the two configurations are identical: 15/20 on short factual questions and 89.5\% ordered-atom recall on long-form tasks in both, with identical text throughout; determinism and batch-composition invariance (a request's output must not depend on its batch-mates) hold 10/10 in every configuration tested, the 2-stage NPU pipeline included; a 2-stage CPU capture of the same model is the one 9/10, differing by a single rewritten contraction. Long free-form generation is not guaranteed bit-identical, because the packed variant is a different compiled graph and greedy decode occasionally flips a near-tied argmax; two of ten 128-token generations differ by one immediately re-converging token, and the 2-stage NPU pipeline lands at the same eight of ten against its own unpacked baseline. On a more shape-sensitive model the divergence is larger, but two packed builds differing only in slot count agree with each other less often than either agrees with the unpacked baseline, which places the sensitivity in compiled-graph shape rather than in packing itself. Single-stage packed isolation is verified on five models across three families (Llama, Qwen, and Phi, up to 14B-class), including eight concurrent identical prompts decoding byte-identically. The structural limits are those of a static path: the window partition is fixed and uniform rather than paged, so a lightly-used slot still reserves its full region; a prompt must fit its slot's region outright and is refused at admission when it cannot, while a generation that outgrows the region slides within it under the pinned-sink rule of \S\ref{sec:npu}; the slot count is baked into the IR shape, so changing it means re-exporting and restarting every stage; decoding remains greedy-only. Within those limits the picture of \S\ref{sec:npu} changes. The placement solver never chose the NPU because single-stream decode there costs about $4\times$ the iGPU per token; packing divides that cost by the concurrent slot count. The 0.2--0.3~ms marginal row cost would also make a $K$-token speculative verify nearly free on this path, a composition we have not yet built.

\subsection{Limitations}

The system assumes a trusted, reliable network. There is no fault tolerance, authentication, or encryption. WiFi latency varies by $\pm$2~ms (averaging out over many tokens). All nodes must run identical software stacks (same OpenVINO, same Python). Every benchmark reported is scoped to integrated GPUs: CPU-hosted stages work (Table~\ref{tab:gemma_dist}) but same-SoC CPU stages contend with the GPU for memory bandwidth, and while NPU stages are now supported through the separate static, stateless export path of \S\ref{sec:npu} (the NPU plugin cannot compile the dynamic stateful \texttt{ReadValue}/\texttt{Assign} shards the rest of the paper is built on), single-stream NPU decode still costs roughly $4\times$ the iGPU per token, and a same-SoC NPU stage shares LPDDR bandwidth with the iGPU, so it does not raise memory-bandwidth-bound decode throughput. The packed serving mode of \S\ref{sec:packed} improves the NPU's multi-user throughput through concurrency, and covers the same 8B-class two-stage pipeline the results tables are built on, but its graph-level slot-scaling ceiling is still a 1.5B-class measurement, correctness verification reaches 14B-class only single-stage, and it has not been tried on the 70B-class deployment of \S\ref{sec:llama70b}.

\textbf{Power and thermal.} We do not report power draw or thermal throttling measurements. For laptop-class AI PCs running sustained GPU inference, thermal throttling is a real concern---anecdotally, the ASUS Zenbook~S~14 fan runs continuously during benchmarks, and prolonged runs on battery would trigger power-limit throttling. A production deployment would need to account for per-node thermal headroom; we expect this to bound sustained multi-user throughput below the measured peaks but have not quantified the effect.

\textbf{WAN characterization (three methods reported).} The headline sweep in Table~\ref{tab:wan} uses \texttt{time.sleep()} on the worker's recv/send boundaries to inject deterministic one-way delay; Table~\ref{tab:real_wan} adds a release-time-queue TCP proxy on the rainier 3-stage path to capture per-RTT cwnd dynamics; \S\ref{sec:tiber} adds a real cross-subnet WAN measurement on Intel Tiber Cloud over Tailscale's DERP relay. The three methods bracket the full WAN cost: sleep-sim isolates the latency dimension, queue-proxy captures multi-segment cwnd effects, and the Tiber Cloud measurement adds real relay-server queueing on top. The most consequential lesson is that for 501~KB FP32 logits returns, real-WAN cost scales \emph{per segment} not \emph{per RTT} --- which is why top-1 logits compression (\S\ref{sec:topk_compress}) gives an $8.17\times$ speedup on DERP-relayed paths but only $1.12\times$ on LAN. The right WAN cost model depends on the dominant per-segment cost; we recommend reporting all three.

\textbf{External 70B reference.} The 4-stage 70B result (\S\ref{sec:llama70b}) is bit-exact relative to the same-topology target-only baseline, but a monolithic single-graph 70B INT4 reference would be a stronger correctness check. Building one on the same toolchain failed: the FP16 calibration weights ($\sim 141$~GB) plus NNCF compression workspace exceeded the export machine's $133$~GB RAM. A node with $\geq 256$~GB RAM, or a chunked NNCF flow that streams layers through quantization, would unblock this.

\subsection{Negative Results}
\label{sec:negative}

Optimization attempts that failed, documented here because the failure modes are informative:

\textbf{Early exit.} We applied the final model's RMSNorm~+~lm\_head to layer~15 outputs, hoping high-confidence tokens could skip the second half of the network. The lm\_head never produced confident predictions at layer~15 (max softmax probability stayed below 50\%), and the $800$~MB matrix multiply added $80$~ms overhead per token. The fundamental issue: the lm\_head is calibrated for layer-31 representations, not layer-15. A purpose-trained lightweight exit head might work but requires calibration data we did not collect.

\textbf{Asymmetric shard planning.} A 12/20 layer split (vs.\ the default 16/16) yielded 2\% improvement---within measurement noise. The bottleneck is network round-trip cost, not compute imbalance between stages.

\textbf{\texttt{KV\_\brk CACHE\_\brk PRECISION} and \texttt{INFERENCE\_\brk PRECISION\_\brk HINT} on Arc iGPU.} We tested all documented OpenVINO GPU precision hints to see if INT8 KV cache would accelerate the memory-bound decode path. On Arc B390, the INT8 KV configuration runs $4\%$ \emph{slower} than default ($20.86$ vs.\ $21.74$~tok/s on monolithic Llama~3.1~8B INT4)---dequantization overhead exceeds the memory bandwidth savings at this sequence length on this GPU. Output is bit-exact across all hint values. Per-stage asymmetric precision would not help for the same reason (the GPU is already memory-bound-limited, not compute-bound). Recipe-based cache compression such as GEAR~\cite{kang2024gear} targets a different bottleneck: near-lossless 4-bit caches buy \emph{capacity} (longer contexts or more concurrent streams in fixed memory) at the price of additional decompression compute. Our result argues against expecting per-token bandwidth savings from cache quantization on this GPU; it says nothing against capacity-motivated compression, which we have not evaluated.

\textbf{Async / threaded overlap of draft and target.} At LAN, trying to overlap draft drafting on Python threads with target verify (using \texttt{SO\_KEEPALIVE}, \texttt{TCP\_NODELAY}, and GIL-releasing \texttt{infer()} calls) yields no measurable gain: $11.08$ vs.\ $11.05$~tok/s at $K\!=\!10$, $100$~ms/hop simulated. The ``all $K$ drafts accepted'' case where async helps has probability $p^K \approx 0.2\%$ at $K\!=\!10$, so simple speculation of next-step drafts during the target's network wait almost always wastes the speculative work. Tree speculation covering multiple accept-count branches would work but requires a 4D attention mask our OV export does not currently support.

\textbf{C++ port of the spec-decode loop.} We considered rewriting \texttt{spec\_\brk decode\_\brk greedy()} in C++ to reduce Python overhead. A dedicated profile (\texttt{reproduction/\brk scripts/\brk bench/\brk bench\_\brk feed\_\brk overhead.py}) finds Python is $0.6\%$ of wall time ($27$~ms out of $4521$~ms for $K\!=\!3$, 128 tokens); OpenVINO's \texttt{req.infer()} accounts for $99.3\%$. The effort would move nothing. The measured $22\%$ Bad-Speculation rate in VTune P-core analysis is OpenVINO's own runtime dispatch/wait loop, not user Python.

\section{Conclusion}
\label{sec:conclusion}

A fleet of commodity Intel AI PCs can serve multi-user LLM inference at interactive speeds \emph{above} single-user monolithic throughput on the same hardware, can scale to models that no single fleet member can host, and can do so across network links where naïve pipeline parallelism is unusable. The two-node full-stack Llama~3.1~8B configuration---$v_5$\_beam-exported shards, mask-based speculative decoding, two-stream micro-batching---reaches $43.97$~tok/s system throughput ($1.79\times$ the monolithic single-user baseline; ${\sim}4\times$ naïve distributed decode under a simulated $100$~ms/hop WAN). A three-node 8B configuration scales to $64.67$~tok/s 3-stream ($2.64\times$ mono, three concurrent users). A four-node Llama~3.1~70B INT4 deployment over a Tailscale DERP relay reaches $6.43$~tok/s 2-stream ($3.1\times$ over the same-topology target-only baseline, bit-exact), demonstrating that the fleet model extends past 8B to model sizes that genuinely require distribution.

The three composing techniques each address a distinct bottleneck and are individually validated with bit-exact output vs.\ their respective baselines:

\begin{itemize}[leftmargin=*]
\item The \texttt{beam\_idx} Gather injection unlocks the OpenVINO GPU plugin's \texttt{Indirect\brk KVCache} fusion for per-stage exports, producing shards at monolithic parity~(\S\ref{sec:shards}).
\item Mask-based KV-cache rewind sidesteps the $\sim 48$~ms-per-call \texttt{query\_state}/\texttt{set\_state} round-trip that would otherwise make speculative decoding a net loss on stateful OpenVINO~(\S\ref{sec:spec}).
\item Independent-state micro-batching fills the stage-idle windows that $v_5$\_beam fusion leaves smaller---each stream isolated via its own \texttt{compile\_\brk model}~(\S\ref{sec:microbatch})---yielding $1.80\times$ scaling at the 2-stream operating point.
\end{itemize}

Code, export scripts, all raw benchmark logs, and VTune reports are available at \url{https://github.com/labscommunity/pipeline-sharded-inference-paper} (the top-level \texttt{reproduction/} directory contains the full export pipeline, distributed coordinator and worker, benchmark drivers, and per-section reproduce scripts).

\bibliographystyle{plainnat}
\bibliography{references}

@article{borzunov2023petals,
  title={Petals: Collaborative Inference and Fine-tuning of Large Models},
  author={Borzunov, Alexander and Baranchuk, Dmitry and Dettmers, Tim and Ryabinin, Max and Belkada, Younes and Chumachenko, Artem and Samygin, Pavel and Raffel, Colin},
  journal={arXiv preprint arXiv:2209.01188},
  year={2023}
}

@article{tong2025parallax,
  title={Parallax: Efficient {LLM} Inference Service over Decentralized Environment},
  author={Tong, Chris and Jiang, Youhe and Chen, Gufeng and Zhao, Tianyi and Lu, Sibian and Qu, Wenjie and Yang, Eric and Ai, Lynn and Yuan, Binhang},
  journal={arXiv preprint arXiv:2509.26182},
  year={2025}
}

@article{macario2025mdillm,
  title={Model-Distributed Inference for Large Language Models at the Edge},
  author={Macario, Davide and Seferoglu, Hulya and Koyuncu, Erdem},
  journal={arXiv preprint arXiv:2505.18164},
  year={2025}
}

@inproceedings{chen2025ktransformers,
  title={KTransformers: Unleashing the Full Potential of CPU/GPU Hybrid Inference for MoE Models},
  author={Chen, Hongtao and Xie, Weiyu and Zhang, Boxin and Tang, Jingqi and Wang, Jiahao and Dong, Jianwei and Chen, Shaoyuan and Yuan, Ziwei and Lin, Chen and Qiu, Chengyu and Zhu, Yuening and Ou, Qingliang and Liao, Jiaqi and Chen, Xianglin and Ai, Zhiyuan and Wu, Yongwei and Zhang, Mingxing},
  booktitle={Proceedings of the 31st ACM Symposium on Operating Systems Principles (SOSP)},
  year={2025}
}

@inproceedings{agrawal2024sarathi,
  title={Taming Throughput-Latency Tradeoff in {LLM} Inference with {Sarathi-Serve}},
  author={Agrawal, Amey and Kedia, Nitin and Panwar, Ashish and Mohan, Jayashree and Kwatra, Nipun and Gulavani, Bhargav S and Tumanov, Alexey and Ramjee, Ramachandran},
  booktitle={Proceedings of the 18th USENIX Symposium on Operating Systems Design and Implementation (OSDI)},
  year={2024}
}

@inproceedings{yu2022orca,
  title={Orca: A Distributed Serving System for Transformer-Based Generative Models},
  author={Yu, Gyeong-In and Jeong, Joo Seong and Kim, Geon-Woo and Kim, Soojeong and Chun, Byung-Gon},
  booktitle={Proceedings of the 16th USENIX Symposium on Operating Systems Design and Implementation (OSDI)},
  year={2022}
}

@inproceedings{xiao2024streamingllm,
  title={Efficient Streaming Language Models with Attention Sinks},
  author={Xiao, Guangxuan and Tian, Yuandong and Chen, Beidi and Han, Song and Lewis, Mike},
  booktitle={International Conference on Learning Representations (ICLR)},
  year={2024}
}

@inproceedings{zhang2025tdpipe,
  title={{TD-Pipe}: Temporally-Disaggregated Pipeline Parallelism Architecture for High-Throughput {LLM} Inference},
  author={Zhang, Hongbin and Wei, Taosheng and Zheng, Zhenyi and Du, Jiangsu and Chen, Zhiguang and Lu, Yutong},
  booktitle={Proceedings of the 54th International Conference on Parallel Processing (ICPP)},
  year={2025}
}

@inproceedings{patel2024splitwise,
  title={Splitwise: Efficient Generative {LLM} Inference Using Phase Splitting},
  author={Patel, Pratyush and Choukse, Esha and Zhang, Chaojie and Shah, Aashaka and Goiri, Inigo and Maleki, Saeed and Bianchini, Ricardo},
  booktitle={Proceedings of the 51st Annual International Symposium on Computer Architecture (ISCA)},
  year={2024}
}

@inproceedings{kwon2023vllm,
  title={Efficient Memory Management for Large Language Model Serving with {PagedAttention}},
  author={Kwon, Woosuk and Li, Zhuohan and Zhuang, Siyuan and Sheng, Ying and Zheng, Lianmin and Yu, Cody Hao and Gonzalez, Joseph E and Zhang, Hao and Stoica, Ion},
  booktitle={Proceedings of the 29th ACM Symposium on Operating Systems Principles (SOSP)},
  year={2023}
}

@article{grattafiori2024llama,
  title={The {Llama 3} Herd of Models},
  author={Grattafiori, Aaron and others},
  journal={arXiv preprint arXiv:2407.21783},
  year={2024}
}

@article{gemma4_2026,
  title={Gemma 4 Technical Report},
  author={{Gemma Team}},
  journal={arXiv preprint arXiv:2607.02770},
  year={2026}
}

@misc{openvino2026,
  title={{OpenVINO} Toolkit},
  author={{Intel Corporation}},
  year={2026},
  howpublished={\url{https://github.com/openvinotoolkit/openvino}},
  note={Open-source toolkit for optimizing and deploying AI inference}
}

@misc{nncf2024,
  title={Neural Network Compression Framework ({NNCF})},
  author={{Intel Corporation}},
  year={2024},
  howpublished={\url{https://github.com/openvinotoolkit/nncf}},
  note={INT4/INT8 weight compression for OpenVINO models}
}

@inproceedings{huang2019gpipe,
  title={{GPipe}: Efficient Training of Giant Neural Networks using Pipeline Parallelism},
  author={Huang, Yanping and Cheng, Youlong and Bapna, Ankur and Firat, Orhan and Chen, Dehao and Chen, Mia and Lee, HyoukJoong and Ngiam, Jiquan and Le, Quoc V and Wu, Yonghui and Chen, Zhifeng},
  booktitle={Advances in Neural Information Processing Systems (NeurIPS)},
  year={2019}
}

@inproceedings{narayanan2019pipedream,
  title={{PipeDream}: Generalized Pipeline Parallelism for {DNN} Training},
  author={Narayanan, Deepak and Harlap, Aaron and Phanishayee, Amar and Seshadri, Vivek and Devanur, Nikhil R and Ganger, Gregory R and Gibbons, Phillip B and Zaharia, Matei},
  booktitle={Proceedings of the 27th ACM Symposium on Operating Systems Principles (SOSP)},
  year={2019}
}

@article{kang2024gear,
  title={{GEAR}: An Efficient {KV} Cache Compression Recipe for Near-Lossless Generative Inference of {LLM}},
  author={Kang, Hao and Zhang, Qingru and Kundu, Souvik and Jeong, Geonhwa and Liu, Zaoxing and Krishna, Tushar and Zhao, Tuo},
  journal={arXiv preprint arXiv:2403.05527},
  year={2024}
}

@article{tian2025skipkv,
  title={{SkipKV}: Selective Skipping of {KV} Generation and Storage for Efficient Inference with Large Reasoning Models},
  author={Tian, Jiayi and Azizi, Seyedarmin and Zhao, Yequan and Potraghloo, Erfan Baghaei and McPherson, Sean and Sridhar, Sharath Nittur and Wang, Zhengyang and Zhang, Zheng and Pedram, Massoud and Kundu, Souvik},
  journal={arXiv preprint arXiv:2512.07993},
  year={2025}
}

@inproceedings{leviathan2023fast,
  title={Fast Inference from Transformers via Speculative Decoding},
  author={Leviathan, Yaniv and Kalman, Matan and Matias, Yossi},
  booktitle={Proceedings of the 40th International Conference on Machine Learning (ICML)},
  year={2023}
}

@article{chen2023accelerating,
  title={Accelerating Large Language Model Decoding with Speculative Sampling},
  author={Chen, Charlie and Borgeaud, Sebastian and Irving, Geoffrey and Lespiau, Jean-Baptiste and Sifre, Laurent and Jumper, John},
  journal={arXiv preprint arXiv:2302.01318},
  year={2023}
}

\end{document}